\documentclass[reprint,12pt]{elsarticle}

\usepackage{amssymb}

\usepackage{graphicx}
\usepackage{xcolor}
\usepackage{amsmath}
\usepackage{amsthm}
\usepackage{bm}
\usepackage{bbm}
\usepackage{mathdots}
\usepackage{lipsum}
\usepackage{verbatim}
\usepackage{nccmath}
\usepackage[caption=false]{subfig}
\newcommand{\T}{\text{\footnotesize\ensuremath{T}}}
\newcommand{\N}{\text{\scriptsize\ensuremath{N}}}
\newcommand{\h}{\text{\tiny\ensuremath{H}}}
\newcommand{\n}{\text{\tiny\ensuremath{N}}}

\biboptions{sort&compress}
\usepackage{amsfonts} 
\DeclareMathSymbol{\shortminus}{\mathbin}{AMSa}{"39}
\journal{Annals of Physics}
\usepackage{hyperref}

\begin{document}

\begin{frontmatter}



	\title{Path Integrals from Spacetime Quantum Actions}

\author[a,b]{N.~L.\ Diaz}
\author[c]{J.~M.\ Matera}
\author[c,d]{R.~Rossignoli}

\address[a]{Information Sciences, Los Alamos National Laboratory, Los Alamos, New Mexico 87545, USA}
\address[b]{Center for Nonlinear Studies (CNLS), Los Alamos National Laboratory, Los Alamos, New Mexico 87545, USA}
\address[c]{Departamento de F\'isica-IFLP/CONICET, Universidad Nacional de La Plata, C.C. 67, La Plata (1900), Argentina}
\address[d]{Comisi\'on de Investigaciones Cient\'{\i}ficas (CIC), La Plata (1900), Argentina}

\begin{abstract} 
The possibility of extending the canonical formulation of
quantum mechanics (QM) 
to a space-time symmetric form
has recently attracted wide interest. 
In this context, a recent proposal has shown that a spacetime symmetric many-body
extension of the Page and Wootters mechanism naturally leads to the 
so-called Quantum Action (QA) 
 operator, a quantum version of the 
action of classical mechanics. 
In this work, we focus on connecting 
the QA with the well-established
Feynman's Path Integral (PI).  
In particular, we present a novel formalism which allows one to identify 
the ``sum over histories''
with a quantum trace, where the role of the classical action is replaced by the 
corresponding QA. 
The trace is defined in the extended Hilbert space resulting from assigning a conventional Hilbert space to each time slice and then 
taking their tensor product. 
The formalism opens the way to the application of quantum computation protocols to the evaluation of PIs and general correlation functions, and reveals that different representations of the PI arise from distinct choices of basis in the evaluation of
 the same trace expression. 
The Hilbert space embedding of the PIs also discloses a new approach to their continuum time limit. Finally, we discuss how the ensuing canonical-like version of QM inherits many properties from the PI formulation, thus allowing an explicitly covariant treatment of spacetime symmetries.
	\end{abstract}



\end{frontmatter}

	\section{Introduction}

        Feynman introduced his Path Integral (PI) formulation \cite{Feynm.1948,feyn.05} as a way to link the Lagrangian formalism with quantum mechanics (QM). It was 
        soon realized that through the 
        classical action, it could provide an explicit 
    spacetime covariant description of quantum systems, able to circumvent the limitations of the canonical Hamiltonian formulation \cite{wei.95}. 
        The formulation became a successful tool to make precise predictions from intuitive (classical) models in high energy and condensed matter physics   \cite{srednicki2007,altland.2010},  being now a typical topic in textbooks. More general forms and representations of PIs also followed 
        \cite{schulman.81}, together with computational applications such as in quantum Monte Carlo techniques \cite{wilson1974confinement, barker1979quantum}.

          Despite the PI success,  there has recently  been an increasing interest in building a spacetime symmetric extension of the quantum canonical formalism itself \cite{dia.19,diaz.21,DMR.24,g.22,fit.15,ho.17,cot.18,ajp.23}, as well as  in  quantum time treatments \cite{QT.15,di.19,mc.13,b.16,m.17,nik.18,b.18,m.18,cas.18,bru.19,p.19,col.19,we.20,lm.20,pla.20,fav.20,fot.21,pa.21,ca.20,he.20,ch.22,LBCR.22,DBLMRC.23, alonso.24}. 
          Besides the important connections between these approaches and quantum gravity \cite{PaW.83,ish.94,br.96,Ish.98,Ma.97,g.09,Ku.11,boj.11,chat.20,ho.21}, including  the emergence of spacetime \cite{ryu.06, van.10,sw.12,ca.17},
          the  need for such symmetry arises naturally within the field of quantum information whose inherent setting is a Hilbert space approach: the spacetime asymmetries constitute a fundamental obstacle for the development of genuine spacetime extensions of quantum information related insights \cite{dia.19,diaz.21,g.22,fit.15,ho.17,cot.18,b.16,ca.17,cos.18,mil.21}. 
          Being based on a ``sum over histories'' rather than a Hilbert space, the  standard PI formulation is not directly able to overcome this problem. This is put in 
          evidence  by the open debate about the  possibility of a useful definition ``spacetime states'' in QM (composite states across time)  \cite{ish.94,fit.15,ho.17,cot.18,dia.19,diaz.21, g.22,ajp.23, DMR.24}, a discussion which transcends the PI's grasp. 
        
        Yet, one may conjecture that fundamental insights, relevant to the construction of the new extensions of QM, and at the same time to the PI formulation, could be obtained if           it were possible to understand the PIs themselves as emerging elements of the extended perspective.           Such possibility should be a reasonable feature           of any fully established spacetime symmetric formulation of QM.           This novel course of action determines the main goal of the present work: the embedding of the PIs in a framework in which the sum over histories acquires Hilbert space meaning.

        For this purpose we will employ an extended Hilbert-space and a concomitant Quantum Action (QA) operator recently introduced in \cite{diaz.21},           as a ``second quantized'' generalization          of the Page and Wootters formalism \cite{PaW.83}. Within this space, the sum over histories can be easily identified with          a quantum trace involving the QA operator. This step shows that the two perspectives, namely the well-known          Feynman's approach and the recent spacetime extensions of canonical QM         (at least in the present form),         can be successfully unified, with important implications to both.  
        
         In particular, we show how distinct representations of the PI now correspond to different bases in the evaluation of the same trace expression. These traces represent correlation functions and inherit the geometrical and spacetime symmetric character of the PI formulation. In addition, these expressions can be applied to any quantum system and computed via conventional quantum computation techniques. A new Hilbert space approach to the continuum case is also developed, which unveils (through an emerging timescale invariance) an equivalence between trace expressions and expectation values in spacetime vacuum states. A direct Hilbert space treatment of spacetime symmetries becomes also feasible.  A preliminary version of this approach to the PI was introduced in \cite{diaz.21}, as a by-product of the QA. A related discrete approach in the context of variational methods for many-body fermion systems was also recently introduced in \cite{CM.21}, showing the potential of this viewpoint for practical and numerical applications. 
                  
         The formalism is first developed in section \ref{sec:II} in its time-sliced version. In this scenario, the extended expressions for correlation functions and the connection between extended bases and PI representations are also introduced, including an explicit example of the use of non-local in time bases. General remarks regarding finite dimensional systems and computational applications are also presented.
         The continuum formulation is presented in section \ref{sec:continuum}, where the relation with the usual PI approach, the treatment of generating functionals and the application  to  relativistic scenarios are discussed. The appendices contain the technical proofs and additional details, including diagonalization of the QA in the general interacting case.
         Conclusions and perspectives are finally drawn in section \ref{sec:IV}.  
        
\section{Sum over histories as a quantum trace} \label{sec:II}
\subsection{Hilbert space time slicing}
\label{sec:htslicing}

We begin our exposition by considering the common example of PIs describing a single $1d$ particle. All of the ideas can be immediately generalized to general bosonic systems as we point out throughout the section. In section \ref{ApAA} we also remark how our approach is more general as it applies to any quantum mechanical systems, including finite-dimensional ones.

A standard procedure to obtain the Feynman's formulation from the canonical one is to express the propagator as
\begin{equation}\label{eq:prop}
    \langle q'|e^{-iHT}|q\rangle
    =\int       \medop{\prod_{t=1}^{N-1}} dq_t\,\medop{\prod_{t=0}^{N-1}} \langle q_{t+1}|e^{-iH\epsilon}|q_t\rangle
\end{equation}
with $H$ a time-independent Hamiltonian, $q_0=q$, $q_\n=q'$, $\epsilon=\T/\N$ and where we used $\int dq\, |q\rangle \langle q|=\mathbbm{1}$ 
    (we also set $\hbar=1$). Each term in the integrand can then be related to the exponential of the action up to first order in $\epsilon$. On the other hand, since the integrand is a product of matrix elements of $e^{-iH\epsilon}$, it has a natural representation in a new Hilbert space $\mathcal{H}:=\otimes_t \mathfrak{H}_t$ built upon the tensor product
    of $N$ copies of the conventional $\mathfrak{H}$, one for each slice:

	\begin{equation}\label{eq:tslices}
 \begin{split}
   \medop{\prod_{t=0}}^{\N-1}
	\langle q_{t+1}|e^{-iH\epsilon}|q_t\rangle
	   &= \langle q_1 q_2\dots q_\n| 
	   {\otimes}_{t=0}^{N-1}e^{-iH\epsilon}|q_0 q_1\dots q_{{\n}\shortminus 1}\rangle\\
   & =\langle \textbf{q}'| 
	   e^{i\mathcal{P}_t\epsilon}
	 	  {\otimes}_{t=0}^{N-1}
	    e^{-iH\epsilon} 
	   |\textbf{q}\rangle
    \end{split}
	\end{equation}
with $|\textbf{q}\rangle:=|q_0q_1\dots q_{\n\shortminus 1}\rangle=\otimes_t |q_t\rangle$
a basis of quantum states to which we may refer as \emph{quantum trajectory states}.

In the last equality  we have changed the ordering of $\langle q_1\dots q_\n|$ to $\langle \textbf{q}'|=\langle q_\n q_1\dots q_{\n-1}|$ such that both the ket $|q\rangle$ and the bra $\langle q'|$ appear in the Hilbert $\mathfrak{H}_0$ (which may be identified with $\mathfrak{H}_\n$). This was implemented through the application of a unitary ``time translation operator'' 
defined by
	\begin{equation}\label{eq:Pt}
	   e^{i\mathcal{P}_t\epsilon}|q_1 q_2\dots q_\n
	   \rangle:=|q_\n q_1\dots  q_{
	  \n\shortminus1}\rangle\,.
	\end{equation} 
 This operator translates ``geometrically'' the different Hilbert space time slices, and is unrelated to the dynamical information provided by the Hamltonian.
 As a result, it has naturally emerged  from Eq.\  
	(\ref{eq:tslices}) the dimensionless
 \emph{quantum operator} $\mathcal{S}$ satisfying 
	\begin{equation}\label{eq:S}
e^{i\mathcal{S}}:=e^{i\mathcal{P}_t\epsilon}\otimes_{t=0}^{N-1} e^{-iH\epsilon}\,.
	\end{equation}
	It is natural to denote $\mathcal{S}$ as quantum action (QA): 
integrating \eqref{eq:tslices} in the variables $q_t$ yields (see Eq.\ \eqref{eq:prop}) the \emph{exact} result

\begin{subequations}\label{eq:PI}
	\begin{align}
	    \langle q'|e^{-iHT}|q\rangle
    &= \int \medop{\prod_{t=1}^{N-1}}dq_t\, \langle \textbf{q}'|e^{i\mathcal{S}}|\textbf{q} \rangle\label{eq:PIa} \\
        &={\rm Tr}_{\mathcal{H}}
    \,\left[e^{i\mathcal{S}}|q\rangle_0\langle q'|\right]\,, \label{eq:PIb}
	\end{align}
	\end{subequations}
 where ${\rm Tr}_{\mathcal{H}}$ denotes the trace in the extended Hilbert space and $|q\rangle_0\langle q'|=|q\rangle_0\langle q'|\otimes_{t\neq 0}\mathbbm{1}_t$. We see that the contribution of a single (discrete) path is the matrix element of 
 the operator $e^{i\mathcal{S}}$ associated with the path in question. Thus the matrix elements of the QA are taking the role of the classical action in the conventional PI formulation.
Moreover, while
 equation   \eqref{eq:PIa} is a classical sum over histories, it
represents a particular evaluation of the quantum trace in \eqref{eq:PIb} which employs the quantum trajectory basis $|\bm{q}\rangle$. This can be seen by inserting the completeness relation $\int \prod_{t=0}^{N-1}dq_t|\bm{q}\rangle\langle\bm{q}|=\mathbbm{1}$ in \eqref{eq:PIb}. 

In order to make direct contact with Feynman's formulation 
let us consider a standard Hamiltonian $H=p^2/2m+V(q)$. In this case the left hand side (l.h.s.)  in \eqref{eq:PIa} can be expressed as  the well-known Feynman PI \cite{Feynm.1948}, implying  
	\begin{align}\label{eq:PIF}
	   	   \int_{\,_{q(0)=q}^{q(T)=q'}} 
	   {\cal D}q(t) e^{ iS_{\rm cl} }
	    =\int \medop{\prod_{t=1}^{N-1}}dq_t\, \langle \textbf{q}'|e^{i\mathcal{S}}|\textbf{q} \rangle={\rm Tr}_{\mathcal{H}}
    \,[e^{i\mathcal{S}}|q\rangle_0\langle q'|]\,, 	\end{align}
 where  $S_{\rm cl}$ denotes  the classical action evaluated along the path.  For large $N$ the integrand in \eqref{eq:PIa} must then become proportional to  $e^{i S_{\rm cl}}$ with 
 $\prod_{t=1}^{\N-1}dq_t\propto {\mathcal D}q(t)$. 
 On the other hand, Eq.\ (\ref{eq:PIF}) holds exactly $\forall$ $N\geq 2$ with no classical interpolation between $q_t$, $q_{t+1}$, meaning that in general the matrix elements of $\mathcal{S}$ differ from $S_{\rm cl}$.

In order to show explicitly the relationship between the QA and the classical one let us notice first that
the definition (\ref{eq:Pt}) implies 
\begin{equation}\label{eq:legendre}
    \langle \textbf{q}|e^{i\mathcal{P}_t\epsilon}|\textbf{p}\rangle=e^{i\epsilon \sum_t p_t\frac{(q_{t+1}-q_{t})}{\epsilon}} 
    \langle \textbf{q}|\textbf{p}\rangle\,,
\end{equation}
where $q_\n=q_0$, which reveals a clear 
connection between the matrix elements of $\mathcal{P}_t$, the generator of time translations, and a discrete version of the classical \emph{Legendre transform}. This follows straightforwardly from the canonical relation $\langle q|p\rangle=e^{ipq}/\sqrt{2\pi}$ here applied to $|\textbf{p}\rangle=\otimes_t |p_t\rangle$ and $|\textbf{q}\rangle$, which yields $\langle \textbf{q}|\textbf{p}\rangle
=e^{i\sum_t p_t q_t}/(2\pi)^{N/2}\,.$ Note also that the states $|\textbf{q}\rangle$, $|\textbf{p}\rangle$ are eigenstates of operators $q_t,p_t$ acting on $\mathfrak{H}_t$ and globally satisfying the ``extended'' (but canonical) algebra $[q_t,p_{t'}]=i\delta_{tt'}$ which may be used to define $\mathcal{H}$. 

For a free particle with Hamiltonian $H=H(p)$ equation (\ref{eq:legendre}) is exactly generalized to 	${
	    \langle \textbf{q}|e^{i\mathcal{S}}|\textbf{p}\rangle= \exp\{i\sum_t \epsilon[p_t(q_{t+1}-q_{t})/\epsilon -H(p_t)]\}\langle \bm{q}|\bm{p}\rangle\,.}
	$
Instead, 
for $H=p^2/2m+V(q)$ one can use a Trotter first-order approximation 
  to obtain
\begin{align}\label{eq:bars}
    \langle \textbf{q}'|e^{i\mathcal{S}}|\textbf{q}\rangle =  \langle \textbf{q}'|e^{i\bar{\mathcal{S}}}|\textbf{q}\rangle+\mathcal{O}(\epsilon^2) \,,
\end{align}
where we have defined $e^{i\bar{\mathcal{S}}}:=e^{i\mathcal{P}_t\epsilon}\otimes_t e^{-i (p_t^2/2m)\epsilon}e^{-iV(q_t)\epsilon}$. Then, by using the $\textbf{p}$-completeness relation  one obtains
\begin{align}
    \langle \textbf{q}'|e^{i\bar{\mathcal{S}}}|\textbf{q}\rangle &=\frac{1}{2\pi} \int \medop{\prod_{t=1}^{\N-1}} \frac{dp_t}{2\pi}\,\exp[i\epsilon\sum_t (p_t\dot{q}_t-H(p_t,q_t))]\big\vert^{q_\n=q'}_{q_0=q}\nonumber\\\label{eq:saproxq}
    &=\frac{1}{(\!\sqrt{2\pi i \epsilon/m}\,)^\N}\exp[i\epsilon\sum_t (\tfrac{1}{2}m\dot{q}_t^2-V(q_t))]\Big\vert^{q_\n=q'}_{q_0=q}\,,
  \end{align}
   where we used (\ref{eq:legendre}) and with
 $\dot{q}_t:=(q_{t+1}-q_t)/\epsilon$. 
The result (\ref{eq:bars})-(\ref{eq:saproxq}) is the anticipated relation between the matrix elements of $e^{i\mathcal{S}}$ and the classical action. Equation (\ref{eq:saproxq}) shows that the time-sliced 
Feynman's PI is equal to ${\rm Tr}_{\mathcal{H}}
    \,[e^{i\bar{\mathcal{S}}}|q\rangle_0\langle q'|]$.

Let us remark that the previous results are valid for general bosonic systems as their generalization follows 
straightforwardly by extending conventional algebras, namely 
\begin{equation}\label{eq:alg}
  [q_i,p_j]=i\delta_{ij}\longrightarrow  [q_{ti},p_{t'j}]=i\delta_{ij}\delta_{tt'}
\end{equation}
 for $i,j$ arbitrary quantum numbers. For instance, if  $i$ denotes a spatial index, the extended algebra is \emph{symmetric in spacetime} \cite{diaz.21} (see also sec. \ref{IIIC}). 
The case of time-dependent Hamiltonians is also straightforward, and 
follows from replacing in Eqs.\ (\ref{eq:tslices}),(\ref{eq:S}) $\otimes_t e^{-iH_t \epsilon}\to \otimes_t U[(t+1)\epsilon,t\epsilon]$ with Eqs.\ (\ref{eq:PI}, \ref{eq:saproxq}) holding.
The consideration of general intervals of evolution $\T$ and/or propagators evolving an interval $\T'<\T$ is discussed in  \ref{ApA}, while finite dimensional systems are considered in  \ref{ApAA}. 

\subsection{Time-ordered and thermal correlation functions}\label{sec:corrfunc}

The PI formulation provides an elegant geometrical approach to handle correlation functions which is symmetric in space and time. This is in contrast with the conventional Hilbert space approach:  the canonical formulation defines correlators by specifying the time values of operators in the Heisenberg picture, while the positions of operators in space is usually associated with ``sites'' (and hence different Hilbert spaces). Instead, the PI version of correlators only involves the insertion of e.g. positions $q$ in certain spacetime points. In this section, we show how one can develop a similar  spacetime symmetric treatment within the extended Hilbert space.

Consider the  \emph{tensor product of time evolution operators}
\begin{equation}
    \mathcal{V}^\dag:=\otimes_{t=0}^{N-1} U_t(t\epsilon)=\hat{T}'\exp\Big[-i\sum_{t=0}^{\N-1}\int^{t\epsilon}_0dt'\,H_t(t')\Big]\,,
\end{equation}
 which is separable in time and unitary ($\hat{T}'$ denotes time-ordering in the variable $t'$).
Its action on a tensor product of general operators yields
\begin{equation}\label{eq:ophe}
    \mathcal{V}\left(\otimes_tO^{(t)}\right) \mathcal{V}^\dag=\otimes_t O^{(t)}_H(t)
\end{equation}
 with $O^{(t)}_H(t)$ the evolved Heisenberg operator ``$t$'' at time $t$. Note that the site index is dictating the amount of evolution.
  
Remarkably, 
 $\mathcal{V}$ relates
 $\mathcal{P}_t$ with $\mathcal{S}$ as follows (see proof in the \ref{ApB}):
\begin{equation}\label{eq:unitarytrans}
    e^{i\mathcal{S}}=U_0(\T)\mathcal{V}^\dag e^{i\mathcal{P}_t\epsilon}\mathcal{V}\,.
\end{equation}
 This expression can be used to relate operators $\mathcal{S}$ of different theories as well. For periodic evolution $U_0(\T)=\mathbbm{1}$  and the unitary relation discussed in \cite{diaz.21} is recovered. In addition, (\ref{eq:unitarytrans}) may be extended to consider non separable in time interactions defined by couplings between different time-slices, a physical possibility which lies beyond the reach of conventional QM.

\begin{figure}[t]
\centering
\includegraphics[width=0.7\textwidth]{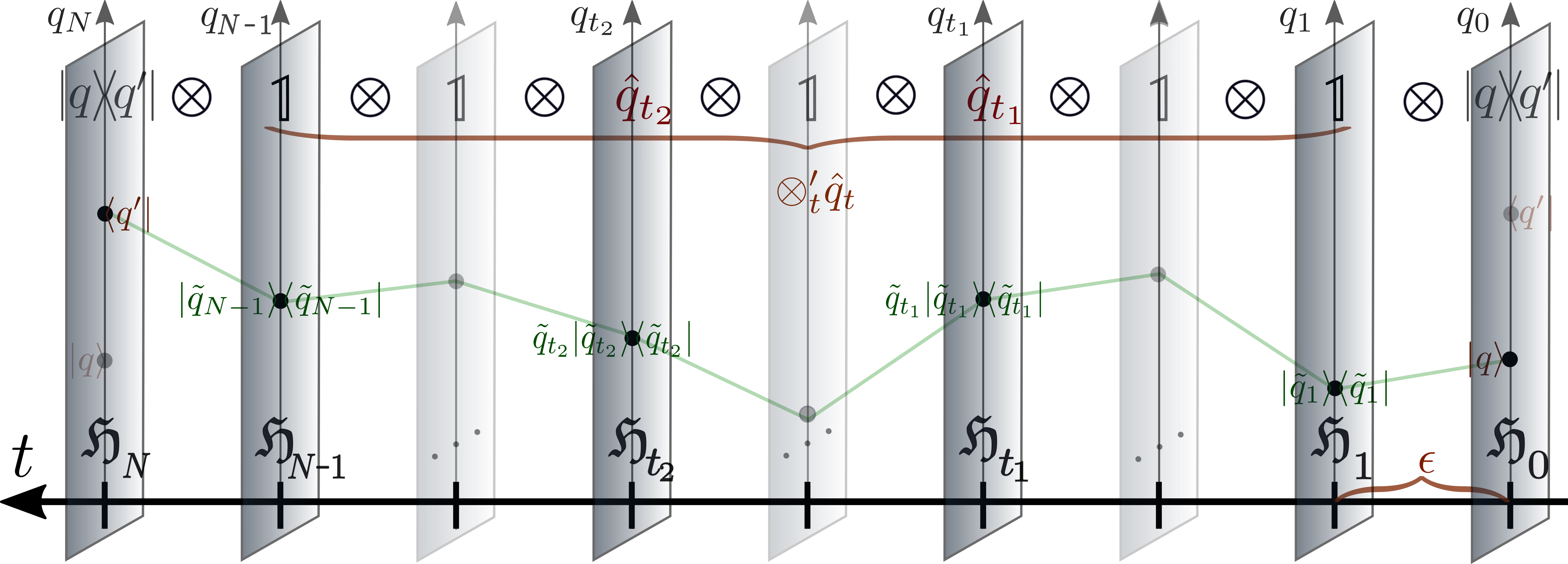}
\caption{Pictorial representation of the operators involved in the extended representation of $\langle q',\T|\hat{T}[q_\h(t_1)q_\h(t_2)]|q\rangle$ for $t_2>t_1$. A proper operator has been ``inserted'' in each Hilbert $\mathfrak{H}_t$. A contribution from a discrete trajectory $\tilde{q}(t)$ is also depicted in order to emphasize the resemblance between the usual interpretation and the Hilbert space construction.}
\label{fig1}
\end{figure}

The result (\ref{eq:unitarytrans}) is particularly useful because it allows the introduction of time evolution via \eqref{eq:ophe} into relations where the time translation operator is involved. In particular, it provides a general expression for \emph{time-ordered correlation functions}, as shown in the \ref{ApB}. For the $1d$  particle of section \ref{sec:htslicing}, this reads:
\begin{align}\label{eq:correl}
\langle q',\T|\hat{T}[q_{\h}(t_1)...\, q_{\h}(t_n)]|q\rangle
&=\medop{\int} \medop{\prod_{t=1}^{\N-1}}\!dq_t\, q_{t_1}...\, q_{t_n} \langle \textbf{q}'|e^{i\mathcal{S}}|\textbf{q} \rangle \nonumber\\&={\rm Tr}_{\mathcal H}\,[e^{i\mathcal{S}} (\otimes'_{t}q_t) |q\rangle_0\langle q'|]\,,
\end{align}
with $|q,\T\rangle=U^\dag(\T)|q\rangle$ and $\otimes'_t$ indicating that only operators on times $t_1,\dots t_n$ are included (and identities otherwise)  such that $\otimes_t '\hat{q}_t|\textbf{q}\rangle=q_{t_1}...\, q_{t_n}|\textbf{q}\rangle$ (see Fig.\ \ref{fig1}). The evolution of the final state $\langle q'|$ arises from the border term $U_0(\T)$ in (\ref{eq:unitarytrans}) while the \emph{time ordering emerges} from the ordering of the time sites.  
The spacetime  trace 
(\ref{eq:correl}) generalizes Eq.\ (\ref{eq:PI}) and its form reflects the corresponding PI expression. It also shares the same geometrical interpretation of the PI, now holding at the Hilbert space level.

Since the result (\ref{eq:unitarytrans}) is a direct consequence of (\ref{eq:Pt}), it holds for general systems and even if $\mathcal{V}$ is non unitary (with $\mathcal{V}^\dag\to \mathcal{V}^{-1}$), which in particular allows to 
discuss \emph{partition functions} and \emph{thermal correlation functions}.
To obtain the former we note that Eq.\ (\ref{eq:PI}) 
implies, for general $H$, 
\begin{equation}\label{eq:traceev}
    {\rm Tr}_{\mathfrak{H}}[e^{-iH T}]={\rm Tr}_{\mathcal{H}}[e^{i\mathcal{S}}]\,.
\end{equation}
Then, for $-iHT\rightarrow -\beta H$ 
 equation (\ref{eq:traceev}) yields the \emph{partition function} of $H$. Moreover, by using Eqs.\  (\ref{eq:unitarytrans}) and (\ref{eq:matrixel}) we now obtain, 
setting $\langle \ldots\rangle_\beta:={\rm Tr}[e^{-\beta H}\ldots]/{\rm Tr}[e^{-\beta H}]$, 
\begin{equation}\label{eq:thcorrel}
   \langle \hat{T}_{\theta} q(\theta_1)...q(\theta_n)\rangle_\beta= \frac{{\rm Tr}_{\cal H}[e^{i\mathcal{S}}\otimes_{t}' q_{t}]}{{\rm Tr}_{\cal H}[e^{i\mathcal{S}}]}
\end{equation}
for  the ``thermodynamic'' \emph{correlation function of thermal states $\propto e^{-\beta H}$}. 
(here, $\theta=it$). 
Notice that unlike Eq.\ (\ref{eq:correl}) we are not specifying any initial (final) state on the $0$-slice since the information of the thermal state is already encoded within $e^{i\mathcal{S}}$ with $U_0(\T)\to e^{-\beta H}$, $\mathcal{V}^\dag\to \mathcal{V}^{-1}$ in Eq.\ (\ref{eq:unitarytrans}). In fact, the linearity and generality of the trace expressions imply (see also  \ref{ApAA}) that the thermal state itself can be obtained as a partial trace (over all modes except those at $t=0$) of the exponential of the quantum action operator \footnote{Notice that the Wick rotation does not affect $e^{i{\cal P}_t\epsilon}$, such that $e^{i{\cal S}}=e^{i{\cal P}_t\epsilon}\otimes_t e^{-H\epsilon}$ in \eqref{eq:thcorrel}-\eqref{rhoeis}, with $\epsilon=\beta/N$.}: 
  \begin{equation}
     \rho=\frac{e^{-\beta H}}{{\rm Tr}_{\mathfrak H}\,[e^{-\beta H}]}=\frac{{\rm Tr}_{t\neq 0}[e^{i\mathcal{S}}]}{{\rm Tr}_{\mathcal H}\,[e^{i\mathcal{S}}]}\,.\label{rhoeis}
 \end{equation}
 
The extension to the case of additional quantum numbers $q\to q_i$, $p\to p_i$ is straightforward and symmetric in space and time (the extended variables are $q_{ti}, p_{tj}$ according to \eqref{eq:alg}). Let us, in fact, remark that when one is considering equal-time correlators, of e.g. two points, the following holds:
\begin{equation}
    \langle q_i q_j\rangle_\beta=\langle q_{0i}q_{0j}\rangle_\mathcal{H}\;,\;\;\; \langle p_i p_j\rangle_\beta=\langle p_{0i}p_{0j}\rangle_\mathcal{H}\,
\end{equation}
 where $\langle \dots \rangle_\mathcal{H}:={\rm Tr}[e^{i\mathcal{S}}\dots]/{\rm Tr}[e^{i\mathcal{S}}]$ in agreement with \eqref{rhoeis}. In this case, both the l.h.s. and the r.h.s. are correlators in the traditional sense, such as those defining spatial entanglement.
 Instead, when we consider operators at different times the expressions become
\begin{equation}
    \langle \hat{T}_{\theta} q_i(\theta_1)q_j(\theta_n)\rangle_\beta=\langle q_{t_1i}q_{t_2j}\rangle_\mathcal{H}\,,\;\;\; \langle \hat{T}_{\theta} p_i(\theta_1)p_j(\theta_n)\rangle_\beta=\langle p_{t_1i}p_{t_2j}\rangle_\mathcal{H}\,,
\end{equation}
which from the conventional perspective (l.h.s.) are no longer  genuine correlators (e.g. the product of operators in general becomes non-hermitian, even for real-time). Remarkably, 
from the $\mathcal{H}$ perspective (the r.h.s.) nothing has changed and these mean values are still ``timeless'' correlators of hermitian operators. 
This shows that the information on whether a separation between operators is space-like or time-like is contained in the QA itself.

 Note also that the reduced state $\rho_V:={\rm Tr}_{\bar{V}}[\rho]$ where the partial trace is over modes outside a region $V$
 can now be recovered from 
\begin{equation}\label{eq:thcorrel2}
   \rho_V= \frac{{\rm Tr}_{t\neq 0, \bar{V}}[e^{i\mathcal{S}}]}{{\rm Tr}_{\mathcal H}\,[e^{i\mathcal{S}}]}
\end{equation}
which is a space time partial trace \emph{outside the space time region} of interest. For quadratic actions this result follows from the previous correlators alone. As a novelty, the formalism allows one to consider partial traces \emph{over general regions of space-time}. In principle, only those associated to space-like hypersurfaces would correspond to conventional quantum states and real entropies but the partial trace is well-defined in general. Interestingly, recent investigations \cite{har.22, nar.22} on the connections between time-like entanglement and geometry in the context of the dS/CFT correspondence use non-hermitian reduced density matrices (in the conventional Hilbert space $\mathfrak{H}$) and complex entanglement ``entropies'' (essentially since a time-like ``distance'' is imaginary).

\subsection{Extended bases and PI representations} \label{sec:extendedbas}
Since Eqs.\ (\ref{eq:PIb})--(\ref{eq:PIF}) and (\ref{eq:correl})--(\ref{eq:thcorrel}) are expressed in terms of traces, different bases of the present extended $\mathcal{H}$,  can now be employed to compute them.
These different bases are formed by a complete set of extended states 
i.e. states in $\mathcal{H}=\otimes_t \mathfrak{H}_t$. 
They include  separable-in-time bases, such as that formed by the states $|\textbf{q}\rangle$ employed  
in section \ref{sec:htslicing} (which generates the usual ``configuration space PI''), as well as, of course,  {\it entangled-in-time} bases, formed by irreducible linear combinations of product states. 
 
It is now convenient to define annihilation  (and creation) operators at time-site  $t$,  \begin{equation}A_t:=e^{i\phi}(\eta q_t+ip_t/\eta)/\sqrt{2}\end{equation} for $\eta,\phi\in\mathbb{R}$  constant, satisfying $[A_t,A_{t'}^\dag]=\delta_{tt'}$. We will denote their vacuum as $|\Omega\rangle=\otimes_t |0_t\rangle$ which is a separable-in-time trajectory state. General extended states are thus obtained by the application of creation operators onto the vacuum. In particular, for $\eta\equiv e^{i\phi}\equiv 1$, $|\textbf{q}\rangle=\exp[-\frac{1}{2}\sum_t\,A^\dag_t (A^\dag_t-2\sqrt{2}q_t)]|\Omega\rangle$ \cite{diaz.21}, showing again that quantum trajectories are particular (and separable) extended states.
 
We can also employ a separable basis of ``coherent trajectory states''
\begin{equation}\label{eq:coherentst}
    |\boldsymbol{\alpha}\rangle:=\exp{\Big[\,\sum_t \alpha_t A^\dag_t\,\Big]}|\Omega\rangle=
    \otimes_t |\alpha_t\rangle_t
\end{equation}
for $|\alpha\rangle_t$ a conventional (unnormalized) coherent state in $\mathfrak{H}_t$ such that 
$$\int \prod_t \frac{d^2\alpha_t}{\pi}e^{-\sum_t |\alpha_t|^2}|\boldsymbol{\alpha}\rangle \langle \boldsymbol{\alpha}|=\mathbbm{1}_{\mathcal H}$$ and $A_t|\bm{\alpha}\rangle=\alpha_t|\bm{\alpha}\rangle$. The ensuing  integral can be easily related to  discretized \emph{coherent-state PIs} (CSPI) under the usual small $\epsilon$ approximation  $\langle \alpha |e^{-i\epsilon H(a,a^\dag)}|\alpha\rangle\approx e^{-i\epsilon H_N(\alpha,\alpha^\ast)}$ for $H=H_N$ with $H_N=:H_N:$ normal ordered. In fact,
\begin{equation}\label{eq:coherentmv}
    \frac{\langle \boldsymbol{\alpha}|e^{i\mathcal{S}}|\boldsymbol{\alpha}\rangle}{\langle \boldsymbol{\alpha}|\boldsymbol{\alpha}\rangle}\approx \exp\big[i\epsilon \sum_t\,(\tfrac{ \alpha^\ast_{t+1}-\alpha^\ast_t}{\epsilon}\alpha_t-H_N(\alpha_t,\alpha^\ast_{t+1}))\big]=e^{i S_{\rm cl}}\,,
\end{equation}
where $S_{\rm cl}$ is the (time-sliced)
classical action for $H_N$ along the path defined by $|\bm{\alpha}\rangle$. 

On the other hand, the non-separable action of the time translation operator suggests the introduction of new \emph{non-local in time basis}: 
we define via Fourier Transform (FT) the operators  \begin{equation}
    A_n:=\tfrac{1}{\sqrt{N}}\sum_t e^{i\omega_n t\epsilon}A_t\label{An}
\end{equation}
with $\omega_n=2\pi n/\T$ for $n \in {\mathbb Z}$, and an ``extended'' vacuum $|\Omega\rangle$ defined by $A_n|\Omega\rangle= A_t|\Omega\rangle=0$. Therefore, we may write \cite{diaz.21}
\begin{equation}\label{eq:normalPt}
    e^{i\mathcal{P}_t \epsilon}=e^{i\epsilon\sum_n \omega_n A^\dag_n A_n}\,,
\end{equation}
which clearly satisfies $e^{i{\cal P}_t\epsilon}O_t e^{-i{\cal P}_t\epsilon}=O_{t+1}$ for $O_t=A_t$, and hence also for any local in time operator as $q_t$ and $p_t$.  
This normal form of $e^{i\mathcal{P}_t\epsilon}$ is invariant under time independent canonical transformations \cite{diaz.21} (and hence independent of the parameter $\eta$ in $A_t$).  And for $\omega_n, \omega_{-n}$ taking symmetric values around $0$ the condition $[\mathcal{P}_t,\sum_t H_t]=0$ (and the invariance of the normal form of $\mathcal{P}_t$) is also verified.

The same coherent state (\ref{eq:coherentst}) can be rewritten in the Fourier basis as $|\boldsymbol{\alpha}\rangle=e^{\,\sum_n \alpha_n A^\dag_n}|\Omega\rangle$ for $\alpha_n:=\sum_t e^{i\omega_n t\epsilon}\alpha_t/\sqrt{\N}$. If we use Eq.\ (\ref{eq:normalPt}) and evalute the trace (\ref{eq:traceev}) in this basis, the Matsubara-like expansion of the coherent state PI is obtained \cite{Fradk.21} (the frequencies $\omega_n$ in $\mathcal{P}_t$ are precisely the Mastubara frequencies). Since the CSPIs here arise from the bases (\ref{eq:coherentst}) we see that the \emph{Matsubara like expansions} in the space of classical functions \emph{correspond to a non-local in time change of basis} in $\mathcal{H}$. 

The Fourier modes also provide a different expansion for the QAs. In particular, for a harmonic oscillator of mass $m$ and frequency $\omega$, 
$$H_t=
\frac{p_t^2}{2m}+\frac{1}{2}m\omega^2q_t^2=\omega \left(A^\dag_t A_t+\frac{1}{2}\right)\,,$$ 
they enable a direct evaluation of the trace in the basis of eigenstates of ${\mathcal S}$: by using $\sum_t A^\dag_t A_t=\sum_n A^\dag_n A_n$ (and $\eta=\sqrt{m\omega}$) we may directly write the QA operator in the normal form   
\begin{equation}\label{eq:dqaction}
\mathcal{S}=\epsilon\sum_n [(\omega_n-\omega)A^\dag_nA_n-\omega/2]\,.
\end{equation}
We see that the QA is diagonal in the \emph{nonlocal-in-time} Fock basis  $|\tilde{\textbf{n}}\rangle=\prod_n [(A^\dag_n)^{\tilde{n}_n}/\sqrt{\tilde{n}_n!}]|\Omega\rangle$ satisfying $A^\dag_n A_n|\tilde{\textbf{n}}\rangle=\tilde{n}_n|\tilde{\textbf{n}}\rangle$.  By using this Fock basis to compute the trace we obtain

\begin{eqnarray} {\rm Tr}\,e^{i{\cal S}}&=&\sum_{\tilde{\textbf{n}}}\langle \tilde{\textbf{n}}|e^{i\mathcal{S}}|\tilde{\textbf{n}}\rangle=e^{-i\omega \T/2}\prod_n \frac{1}{1-e^{i\epsilon (\omega_n-\omega)}}
\label{eq:tra}\;\;\;
=\frac{1}{2i \sin(\omega \T /2)}\,.\end{eqnarray}
One immediately recognizes the ``partition function'' of the harmonic oscillator, in agreement with Eq.\ (\ref{eq:traceev}) (see the proof of \eqref{eq:tra} in  \ref{ApC}).

On the other hand, since the QA is a bosonic  quadratic operator it holds that
\begin{equation}\label{eq:trdet}
    {\rm Tr}\,e^{i{\cal S}}=e^{-i\omega T}{\det}^{-1}[\mathbbm{1}-e^{i S}]\,,
\end{equation}
where the matrix $S$ is defined by
\begin{equation}[{\cal S},A^\dag_t]=\sum_{t'}S_{t't}A^\dag_{t'}\,.
\label{ConmS}\end{equation}
This allows to write the QA as ${\cal  S}=\sum_{t',t}A^\dag_{t'} S_{t't} A_{t}$, which, when we compare it with \eqref{eq:dqaction} yields
$$S_{t't}=\frac{\epsilon}{N}\sum_n (\omega_n-\omega)e^{i\omega_n \epsilon(t-t')}   =\epsilon\Big(i\frac{d}{dt'}-\omega\Big)\delta_{tt'}\,.$$
It is then clear that the product in Eq.\ (\ref{eq:tra}) is the determinant in Eq.\ (\ref{eq:trdet}), with $\epsilon (\omega_n-\omega)$ the eigenvalues of the matrix $S$.
A similar procedure can be employed in (\ref{eq:PI}) to compute propagators, e.g. $\langle 0,\T|0\rangle=e^{-i\omega\T/2}\det[\bar{M}]^{-1}$ with $\bar{M}$ the 
matrix obtained by removing the first column and row of the larger matrix $M=\mathbbm{1}-e^{i{\cal S}}$.

As another example, changing from the trajectory basis of states   $|\bm{q}\rangle$ to a Fourier basis  $|\tilde{\bm{q}}\rangle$, where 
$\tilde{q}_n=\tfrac{1}{N}\sum_t e^{i\omega_n t\epsilon}q_t=\tilde{q}_{-n}^\dag$,  we obtain the useful PI over the Fourier coefficients $\tilde{q}_n$
\cite{Fradk.21,RCR.98}. While this leads to diagonal actions 
for quadratic Hamiltonians, for interacting theories  this basis also serves as a starting point 
for developing the operator equivalent to typical PI approximations based on neglecting ``fast oscillations'' associated with large $n$. For example, one can easily show that by replacing a given QA operator with the same QA but where only the $\tilde{q}_0$ dependence is retained \footnote{i.e. one replaces the term $\sum_t \epsilon V(q_t)=\sum_t \epsilon V(\sum_n e^{-i\omega_n t \epsilon}\tilde{q}_n)$, which is non-local in $n$ beyond quadratic potentials, with $\sum_t \epsilon V(q_t)\to \beta V(\tilde{q}_0)\equiv \beta V(\tilde{q}_0)\otimes \mathbbm{1}_{n\neq 0}$. Other operators including the conjugated momenta for different $n$ are preserved}, the \emph{classical} partition function is recovered from the trace of the exponential of the (Wick rotated) action. Semiclassical approximations \cite{RCR.98} might be also reformulated at the operator level while new schemes could also emerge in this extended canonical formulation.

On the other hand, more complicated and rich scenarios may be considered, as suggested by the diagonalization of QAs with more complicated Hamiltonians. In general, the diagonalization of a QA operator will be associated with a proper extended basis that does not need to have a direct correspondence with a standard PI representation, nor a direct relation with the diagonalization of a classical action. To illustrate this point, let us consider a Hamiltonian $H$ diagonal in some basis $|n\rangle$ such that $H|n\rangle=\lambda_n |n\rangle$. For example,  $H_K=\omega a^\dag a + \lambda (a^\dag a)^2$ is already diagonal in the Fock basis with $\lambda_n=\omega n+ \lambda n^2$. This Hamiltonian emerges in the presence of non-linear Kerr effects and has been employed to generate 
non-classical states of light  \cite{Ker1,Ker2}.
The associated QA has the form
\begin{equation}\label{eq:actionkerr}
    \mathcal{S}_K=\epsilon\sum_n [(\omega_n-\omega)A^\dag_nA_n-\omega/2]-\epsilon \sum_t \lambda (A_t^\dag A_t)^2\,.
\end{equation}
Notice that while the first term is diagonal in the Fourier basis (and non local in the time basis), the term proportional to $\lambda$ is diagonal in the time basis, but it is highly non-local in terms of Fourier modes ($\sum_t \lambda (A_t^\dag A_t)^2=\lambda \sum_{n_1,n_2,n_3,n_4}\delta_{n_1+n_3-n_2-n_4}A^\dag_{n_1}A_{n_2}A^\dag_{n_3}A_{n_4}$). 
This means the QA, which is an hermitian operator, has an eigenbasis which is generally neither ``$t$'' nor ``$n$'' local. This can be seen explicitly by considering the example of two time slices 
 in which case the states  $|k;n_0n_1\rangle=(|n_0n_1\rangle+(-1)^k |n_1n_0\rangle)/\sqrt{2}$ for $k=0,1$, $n_0<n_1$  and $|nn\rangle$  provide an eigenbasis of the QA (this can be checked directly by acting with $e^{i\mathcal{S}_K}$ and using that $|n_0n_1\rangle=\frac{A_{0}^{\dag n_0}}{\sqrt{n_0!}}\frac{A_{1}^{\dag n_1}}{\sqrt{n_1!}}|\Omega\rangle$ satisfies $(H_K)_t  |n_0n_1\rangle=\lambda_{n_t}|n_0n_1\rangle$). Notice that the states $|k;n_0n_1\rangle$ are highly entangled states from the time-mode point of view. In particular, $|k;0n_1\rangle$ are the well-known NOON states 
 \cite{Noon1,Noon2} (with space replaced by time), which have been applied to quantum metrology and sensing providing quantum advantage. 
In general, due to their entangled-in-time  nature of the quantum action eigenstates, as discussed in more detail in \ref{Apkerr}, there is in principle no standard PI representation associated to this basis.  
 Furthermore, these eigenstate states cannot be written in general as permanents, 
 meaning that even if the original basis diagonalizing $H_K$ was a Fock basis, the diagonalization of $\mathcal{S}_K$ involves more complicated states. This is shown explicitly in the  \ref{Apkerr} where we also discuss the general diagonalization of the QA operator for time-independent Hamiltonians.

\subsection{General systems and quantum computational considerations \label{ApAA}}
\begin{figure}[!htbp]
\centering
\includegraphics[width=0.85\textwidth]{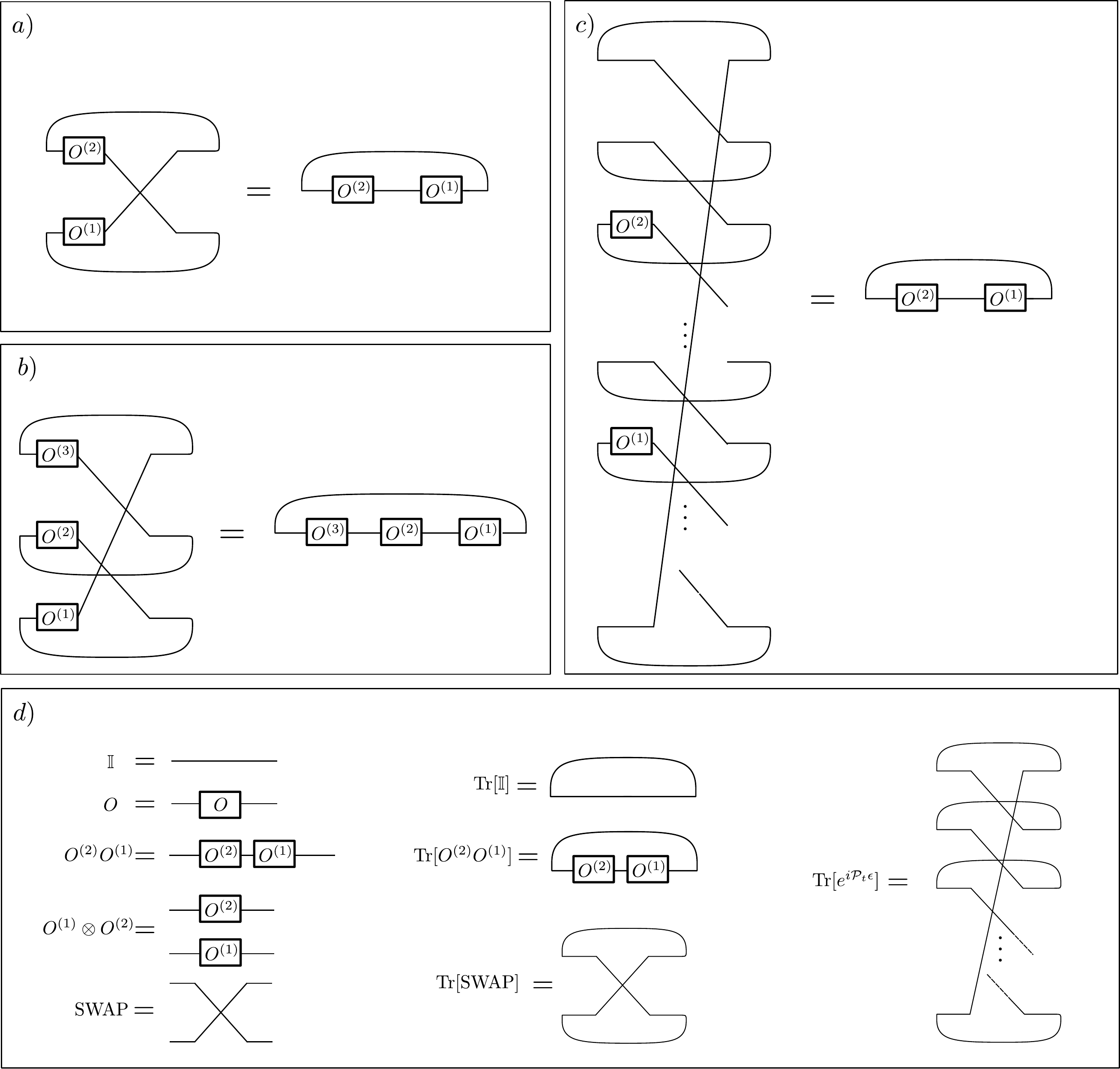}
\caption{\textbf{Diagrammatic representation of the map connecting traces in the extended version of QM to the conventional ones.} 
The map which connects traces of operators within $\mathcal{H}=\otimes_t \mathfrak{h}_t$ with traces of operators in $\mathfrak{h}$ is here represented in tensor network notation (the conventions are depicted in $d)$). In a) we show essentially a SWAP test. In b) a generalization from two to three ``slices''. Instead in $c)$ we show the example of a two-point correlation function, as treated by the formalism. Interestingly, the many sums on the l.h.s. of the diagram, represented by the lines on each vertical row (corresponding each to a different time slice), are precisely the sum over histories. Notice also that the diagrammatic expansion shows that $e^{i\mathcal{P}_t\epsilon}$ can be represented as the composition of many SWAPs.}
\label{figtensor}
\end{figure}

  Regarding general quantum systems, the application of the main ideas are straightforward and non necessarily related to PIs: the key idea is that there is a natural connection between the inner product in a conventional Hilbert space $\mathfrak{H}$ and the inner product in $\mathcal{H}=\mathfrak{H}^{\otimes \n}$. In complete generality it can be expressed for a general basis as follows:
\begin{equation}\label{eq:matrixel}
\begin{split}
  \!\!\!\!{\rm Tr}_{\mathfrak{ H}}\,[O^{(N-1)}\ldots O^{(0)}|i\rangle\langle i'|]&= \langle i'|O^{(N-1)}\ldots O^{(0)}|i\rangle=\sum_{i_1,...,i_{\n-1}} \prod_t\langle i_{t+1}|O^{(t)}|i_t\rangle\\
  &={\rm Tr}_{\cal H}\,[e^{i \mathcal{P}_t\epsilon}\otimes_t O^{(t)} |i\rangle_0\langle  i'|]  
\end{split}
\end{equation}
for $\sum_i |i\rangle \langle i|=\mathbbm{1}$ and $e^{i\mathcal{P}_t \epsilon}|i_0i_1...i_{\n-1}\rangle=|i_{\n-1}i_0i_1...\rangle$. For instance, in the most basic case, 
\begin{align}
{\rm Tr}_{\cal H}[e^{i{\cal P}_t\epsilon} (|i\rangle\langle i'|\otimes \mathbbm{1})]&={\rm Tr}_{\cal H}
\sum_{j} e^{i{\cal P}_t\epsilon}|ij\rangle\langle i'j|=\sum_j\langle i'j|ji\rangle\nonumber\\&=\sum_j \langle i'|j\rangle\langle j|i\rangle=\langle i'|i\rangle\,,\end{align}
which clearly shows the coincidence of the trace in ${\cal H}=\mathfrak H\otimes\mathfrak H$ with the Feynman sum over one intermediate amplitude \cite{Feynm.1948}. It is interesting to notice that for these ``two slices'' the previous works essentially as the SWAP test \cite{buhrman2001quantum}. We can also understand the general case as a generalization of it, with the generator of time translations built from the composition of SWAP operators. This has a nice diagrammatic representation in the language of tensor networks as shown in figure \ref{figtensor}.

Even when no classical notion of trajectory is present, we may still associate the index $t$ with time slices and refer to the states $|\textbf{i}\rangle:=\otimes_t |i_t\rangle$ as quantum trajectories in analogy with $|\textbf{q}\rangle$ (when considering time evolved operators on the l.h.s. a time-ordering will also emerge). 
In other words, we can always establish a \emph{map} between a version of QM which applies a tensor product structure in time and the conventional formulation.
This connection was also employed in \cite{ish.94} to probe theorems related to decoherence functionals \cite{grif.84,gellm.19} within  ``duplicated'' Hilbert spaces of the form $\mathcal{H}\otimes \mathcal{H}$ (we are interested in $\mathcal{H}$ itself and PIs).

  A basic consequence of Eq.\ (\ref{eq:matrixel}) and the linearity of the trace is an expression for \emph{mean values}:
  \begin{equation}\label{eq:meanval}
      {\rm Tr}[O^{(1)}O^{(2)}...\rho ]={\rm Tr}[e^{i \mathcal{P}_t\epsilon}(\otimes_t O^{(t)}) \rho^{(0)}]
  \end{equation}
for $\rho$ a general density matrix in $\mathfrak{H}$ and $\rho^{(0)}$ the same operator acting on the initial slice of $\mathcal{H}$.
Moreover,  the standard definition of partial trace (\ref{eq:matrixel}) implies 
\begin{equation}
     O^{(1)}O^{(2)}...={\rm Tr}_{t\neq 0}[e^{i \mathcal{P}_t \epsilon}\otimes_t O^{(t)}]
\end{equation}
yielding in particular a ``partial trace in time'' for states:

\begin{equation}
    \rho={\rm Tr}_{t\neq 0}[e^{i \mathcal{P}_t \epsilon}\rho^{(0)}]\,.
\end{equation}
Instead, for $O^{(t)}=e^{-iH\epsilon}$ we obtain an expression for the time evolution operator:
$e^{-iH T}={\rm Tr}_{t\neq 0}[e^{i\mathcal{S}}]$.

\begin{figure}[t]
\centering
\includegraphics[width=1\textwidth]{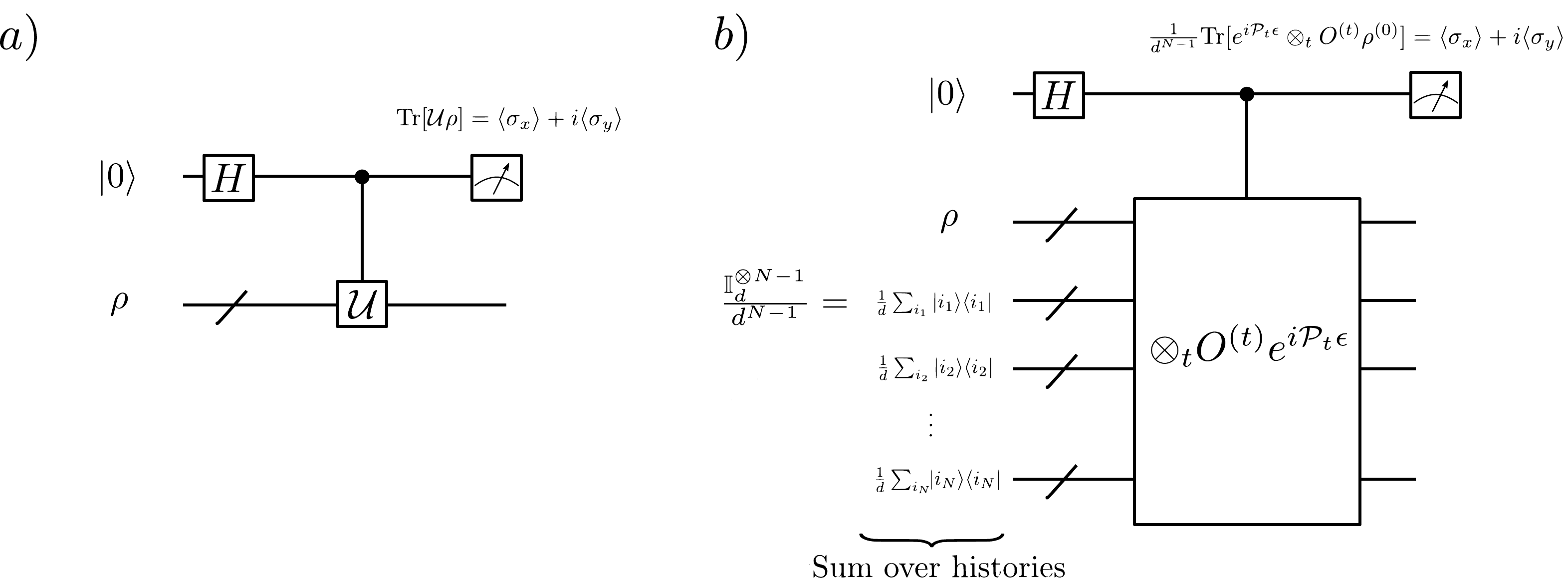}
\caption{\textbf{Protocol for evaluation of PIs via Hadamard test}. $a)$ A generic Hadamard test. The trace ${\rm Tr}\,[{\cal U}\rho]$, where $\rho$ is a an arbitrary (pure or mixed) state of $n$ qubits and ${\cal U}$ an arbitrary unitary operator on $n$ qubits (hence  involving $2^n\times 2^n$ matrix representations) can be evaluated by measuring the averages $\langle\sigma_x\rangle$ and $\langle \sigma_y\rangle$ of the auxiliary top qubit, initially in an eigenstate of $\sigma_z$. The quantum circuit involves just  a Hadamard ($H$)  and a controlled ${\cal U}$ gates.  
$b)$ The application of the protocol to the r.h.s.\ of Eq.\ (\ref{eq:meanval}). When applied to PIs $({\cal U}\rightarrow e^{i{\cal S}}$, Eq.\ (\ref{eq:S}))  the ``sum over histories'' is implicit on the full mixed states at the entry. One may also employ a subset of states, covering a subset of trajectories. 
 Thermal correlation functions and/or ``imaginary time evolution'' (non-implementable through unitary gates) can be computed by replacing $\rho$ and the maximally mixed states $\mathbbm{1}/d$ on the left by suitable thermal states (see sec.\ \ref{sec:corrfunc}). }
\label{fig2}
\end{figure}

 Since tensor products are one basic feature underlying quantum computation \cite{nie.01}, the ability to describe quantum mechanical temporal and thermal properties through an associated ``Hilbert-space-slicing'' is an interesting computational fact by itself. Furthermore, since the main results are expressed in terms of traces of operators, sub-models of quantum computation employing the power of one qubit \cite{DQC1,DQC12} (see Fig.\ \ref{fig2} a)) can be applied to the extended space. As an example, we show on Fig.\  \ref{fig2} a circuit computing the l.h.s. of Eq.\ (\ref{eq:meanval}) for unitary operators $O^{(i)}$, through the ``sum over histories'' implicit in the r.h.s. These tools could also be useful for expanding the discussions on the connection between quantum circuit dynamics and path integrals \cite{pen.17}.

Note that the previous expressions apply directly to both finite distinguishable and bosonic systems.  
The natural extension of the formalism to fermions requires an anticommuting version of the algebra (\ref{eq:alg}) \cite{diaz.21,CM.21}, and is mostly straightforward: in Fock spaces a result such as (\ref{eq:matrixel}) can be easily rewritten in term of Wick's contractions and seen as a direct consequence of the extended algebra.
For relativistic fermions one may employ the sp ``quantum time formalism'' developed in \cite{di.19} for Dirac's theory.

As a final remark, we notice that the postulates of QM \cite{nie.01} are a set of rules assigning physical content to Hilbert space expressions. As a consequence, they in principle  translate directly into the extended ones through the previous relations. Unitary time evolution has been covered throughout the manuscript, while the effects of a measurement may  be in principle introduced by considering the partial trace of the global (system plus measurement apparatus) state undergoing a unitary entangling evolution \cite{nie.01}. More general trace expressions and QAs can be considered in $\mathcal{H}$ (the use of a second order action in section \ref{sec:continuum} is a non trivial example), but their potential physical meaning is left for future investigations.

\section{The continuum time case \label{sec:continuum}}
\subsection{Formalism and small $\tau$ limit \label{sec:consmalltau}}
We will now consider the continuum generalization of the previous ideas. For simplicity, we begin the discussion by returning to the case of a $1d$ particle. 

We need to generalize both the operators and relations such as (\ref{eq:traceev}). In the first case the continuum limit should be obtained by standard means but applied to extended operators: we define \footnote{For convenience of notation, we are using the same variable $t$ in the continuum case to indicate the quantity which in the discrete corresponds to $\epsilon t$ for $t$ the discrete dimensionless index.} $q(t):=q_t/\sqrt{\epsilon}$, $p(t):=p_t/\sqrt{\epsilon}$ such that in the limit $\epsilon\to 0^+$
\begin{equation}\label{eq:algcont}
    [q(t),p(t')]=i\delta(t-t')\,.
\end{equation}
  Similarly $A(t)=A_t/\sqrt{\epsilon}$ implying $[A(t),A^\dag(t')]\to\delta(t-t')$.
Under this limit, the generator of time translations 
is explicitly related
 to the \emph{Legendre transform}:
\begin{equation}\label{eq:ptcontinuo}
    \mathcal{P}_t=\medop{\int} dt\, A^\dag(t)i\dot{A}(t)=\tfrac{1}{2}\medop{\int} dt\, [p(t)\dot{q}(t)-q(t)\dot{p}(t)]
\end{equation}
with $\dot{A}(t)=\sum_n \omega_n e^{-i\omega_n t}A_n/\sqrt{\T}$ and $dt\equiv \epsilon$ \cite{diaz.21}. This is a direct consequence of Eq.\ (\ref{eq:normalPt}). We notice that \eqref{eq:ptcontinuo} is equivalent to $\int dt\, p(t)\dot{q}(t)$. 

On the other hand, the continuum generalization of the ``map'' connecting the inner products of $\mathfrak{H}$ and $\mathcal{H}$ (see e.g. Eqs.\ (\ref{eq:traceev}-\ref{eq:thcorrel})) is non trivial: while the generator of time translations has a proper limit, the concept of translating a ``time-step'' has no longer meaning. Yet, if we introduce an arbitrary time scale $\tau$ the proper  operator translating an amount $\tau$, is well defined as $e^{i\mathcal{P}_t \tau}$ and satisfies  $e^{i\mathcal{P}_t \tau}A(t)e^{-i\mathcal{P}_t \tau}=A(t+\tau)$. 

Analogously, we can define a QA
\begin{equation}\label{eq:tauaction}
    S_{\tau}:=\tau \int dt\,\Big[p(t)\dot{q}(t)-\frac{p^2(t)}{2m}-\mathcal{V}(q(t))\Big]\,
\end{equation}  with  $\mathcal{V}(q(t)):=\tau^{-1}V(q(t)\sqrt{\tau})$ such that for $V(q)$ admitting a power series, $\mathcal{S}_{\tau}$ is at least of order $\sqrt{\tau}$ (up to a constant).
For example, for an harmonic oscillator (for the moment with zero point energy) $\mathcal{S}_\tau= \tau\int dt\,[A^\dag(t)(i\partial_t-\omega)A(t)]$, an operator which shares some similarities (without the time scale $\tau$) with the one proposed in \cite{Sav.99} in the context of Isham's approach \cite{Ish.98} to the continuum-histories formulation of QM \cite{grif.84,gellm.19}.

A priori, it is not evident that the previous definitions are useful.
However, at least heuristically (a more rigorous discussion is provided below and in the next subsection) we can use these ideas to recover PIs in analogy with the discrete case:
using the coherent state basis, we  write
${\rm Tr}[e^{i\mathcal{S}_\tau}]=\int \mathcal{D}^2\alpha(t)e^{-\int dt\,|\alpha(t)|^2}   \langle \alpha(t)|e^{i\mathcal{S}_\tau}|\alpha(t)\rangle
$, with 
\begin{equation}\label{eq:coherentstcont}
    |\alpha(t)\rangle=\exp\left[\medop{\int} dt\,\alpha(t)A^\dag(t)\right]|\Omega\rangle
\end{equation}
 the continuum limit of the state  (\ref{eq:coherentst}) with $\alpha(t) \approx \alpha_t/\sqrt{\epsilon}$. For the harmonic oscillator, 
 the first order in a formal expansion in powers of $\tau$ yields 
\begin{equation} \label{eq:coherentpi}
\begin{split}
      {\rm Tr}\,e^{i\mathcal{S}_\tau}&\approx  \int \mathcal{D}^2\alpha(t)\,e^{i\tau \int dt\,[\alpha^\ast(t)(i\partial_t-\omega)\alpha(t)]}
      ={\det}^{-1}[\shortminus i\tau(i\partial_t-\omega)]\,,
\end{split}
\end{equation}
where we used $\langle \alpha(t)|e^{i \mathcal{S}_\tau}|\alpha(t)\rangle=\langle \alpha(t)|e^{i\omega\tau} \alpha(t+\tau)\rangle$. 
This is precisely what is obtained through the conventional CSPI if one chooses $
\tau$ such that  $\mathcal{D}_F^2\alpha(t)\equiv\mathcal{D}^2[\alpha(t)/\sqrt{\tau}]$, with $\mathcal{D}_F^2\alpha(t)$ the ``Feynman's measure''.
This argument extends to other bases and exposes a simple approach for general theories: in a continuous formulation, after computations such as (\ref{eq:coherentpi}) have been made, the previous identification  yields the conventional PI under changes of variables like $\sqrt{\tau}\alpha(t)\to \alpha(t)$. From this point of view, $\det[i\tau\mathbbm{1}]$ is the  constant conventionally encoded in the measure which ``regularizes'' the divergent functional determinant (\ref{eq:coherentpi}).

Note also that Eq.\ (\ref{eq:coherentpi}) is the formal small $\tau$ limit of Eq.\ (\ref{eq:trdet}) and we are essentially computing the determinant of the ``matrix'' $S$ (with $\epsilon\to \tau$) in Eq.\ (\ref{ConmS}), now the differential operator defined by
\begin{equation}
[\mathcal{S}_\tau,A^\dag(t)]= -i\tau(i\partial_t-\omega)A^\dag(t)\,.
\end{equation}
The formalism  introduces naturally this  differential operator as a linear transformation between \emph{extended} operators $A(t)$, with no reference to a space of classical functions (which arises in some particular evaluations of the trace). 
In addition, higher orders in $\tau$ correct the problems \cite{voon.89,wil.11} associated with the vacuum energy in (continuum) CSPIs. 

From a more rigorous perspective, we can also evaluate the previous trace  in the Fourier basis in which case Eq.\ (\ref{eq:dqaction}) holds (with $\epsilon\to \tau$). Then, in complete analogy with the discrete case (Eq.\ (\ref{eq:tra})), the trace is related to a determinant, which can be easily made finite: under the slight operator modification
\begin{equation}
    \mathcal{S}_\tau=\tau \sum_{n} (\omega_n-\omega+i\tilde{\epsilon} \omega_n^2)A^\dag_n A_n\,,
\end{equation}
 ${\rm Tr}[e^{i\mathcal{S}_\tau}]$ becomes a \emph{finite} determinant for $\tau\tilde{\epsilon}>0$, opening interesting mathematical perspectives (we may also set $\tilde{\epsilon}=\tilde{\epsilon}(\tau)$, see   \ref{ApC} and the conjecture therein). 
 In the following we omit the new term but, where needed, it can be easily restored without compromising the main results. 

\subsection{Generating functionals and $\tau$-invariance}\label{genfunc}

Quite remarkably, 
the operator $\mathcal{S}_\tau$ has important \emph{$\tau$-invariant}
properties 
 which allow a simple definition and evaluation of generating functionals, holding for finite $\tau$.
We will now prove it in a clear example 
by employing exclusively properties of operators and traces in Hilbert space: no subtle mathematical definition of infinite dimensional measures is necessary. Yet, both the simplicity that characterizes PIs and the familiar connection to classical physics can be recovered in the Hilbert space $\mathcal{H}$.

Consider the QA operator $\mathcal{S}_\tau[j]$ which is a function of a current $j(t)$ appearing in the potential energy $V(q)=m\omega^2 q^2/2-\sqrt{m}j(t)q$. We can expand it as
\begin{equation}\label{eq:actionj}
\begin{split}
      \mathcal{S}_\tau[j]&=\tau \int dt\,\Big[A^\dag(t)(i\partial_t-\omega)A(t)+\sqrt{\tfrac{m}{\tau}}j(t)q(t)\Big]\\
   &=\tau \sum_{n=-\infty}^\infty \Big[ (\omega_n-\omega)A^\dag_n A_n+\frac{j_{\shortminus n}}{\sqrt{2\omega \tau}}A_n+\frac{j_n}{\sqrt{2\omega \tau}}A^\dag_n \Big]\,.
\end{split}
\end{equation}
where we employed the nonlocal in time Fourier modes $A_n$ and defined $j_n:=\int dt\,e^{i\omega_n t} j(t)/\sqrt{\T}$,  the Fourier coefficients of the source current $j(t)$.
The $n$-representation reveals immediately  an important  \emph{unitary relation} between quantum actions with and without a source:  
 \begin{equation}\label{eq:unitaryrel}
    \mathcal{S}_{\tau}[j]=\mathcal{U}^{\dag}\left( \mathcal{S}_{\tau}[0]+\mathbbm{1}\, S_{\rm cl}^{\star}[j]\right) \mathcal{U}\,,\;\;{\rm for}\;\;\;
    \mathcal{U}^{\dag} A_n \mathcal{U}=A_n+\frac{j_n}{\sqrt{2\omega\tau}(\omega_n-\omega)} 
\end{equation}
    with 
    \begin{equation}\label{eq:scl}
        S_{\rm cl}^{\star}[j]=-\sum_n \frac{|j_n|^2 }{2\omega(\omega_n-\omega)}=\frac{i}{2}\int dt dt'\, j(t')G(t-t')j(t)\,, 
    \end{equation}
    the \emph{classical} action (not an operator) evaluated in the classical solution. 
    Here $G(t-t')$ is the Green function of the differential operator $(\partial_t^2+\omega^2)$ whose Fourier expansion
    $$G(t-t')=i\sum_n \frac{e^{-i\omega_n(t-t')}}{T(\omega_n^2-\omega^2)}\,,$$
    appears naturally in (\ref{eq:unitaryrel}) by employing (\ref{eq:actionj}) and the relation
    $(\omega_n-\omega)^{\shortminus 1}-(\omega_n+\omega)^{\shortminus 1}=2\omega (\omega_n^2-\omega^2)^{\shortminus1}$ (we assume as usual no caustics). 
    This result can also be  seen by  noting that the classical action along an arbitrary trajectory 
    is related to the average $\langle \alpha(t)|{\cal S}_\tau[j]|\alpha(t)\rangle$ of the quantum action in the corresponding trajectory
    coherent state (see \ref{ApD} for the details). 
    
We now define the generating functional for this theory and arbitrary $\tau$ as \begin{equation}\label{eq:generatingf}
    Z_\tau[j]:={\rm Tr}[e^{i\mathcal{S}_\tau[j]}]\,.
    \end{equation}
    For small $\tau$, the considerations made in section \ref{sec:consmalltau} suggest a connection to the usual definition of the generating functional $Z[j]=\int \mathcal{D}q\, e^{i S_{\rm cl}[j]}$, where the classical action depends on $j(t)$ ($S_{\rm cl}[j]=\int dt\,[m\dot{q}^2(t)/2-m\omega^2q^2(t)/2+ \sqrt{m}j(t)q(t)]$).
    Remarkably, since the transformation (\ref{eq:unitaryrel}) preserves the trace, the ratio  $Z_\tau[j]/Z_\tau[0]$ is actually \emph{$\tau$-invariant} and its evaluation immediate:
        \begin{equation}\label{ZZ}
        \frac{Z_\tau[j]}{Z_\tau[0]}=e^{iS^{\star}_{\rm cl}[j]}
    \end{equation}
          in agreement with the standard result $Z[j]=Z[0]e^{iS^\star_{\rm cl}[j]}$ \cite{Fradk.21}.  We remark that Eq.\ \eqref{ZZ} holds exactly $\forall$ $\tau\neq 0$. Let us also mention that a similar invariance holds in the discrete time-slice formulation and can be developed by similar means.

      Moreover, 
    in the limit $\T\to \infty$ (considering a symmetric time interval and the addition of a small imaginary part to $\omega$), 
    $\sum_n 2\pi/\T\to \int d\tilde{\omega}$ implying $G(t-t')\to D_F(t-t')$ and hence
    \begin{equation}\label{eq:genfeyn}
         Z_\tau[j]=Z_\tau[0]\exp\Big(-\tfrac{1}{2}\medop{\int} \!dt dt'\, j(t')D_F(t-t')j(t)\Big)
   \end{equation} 
    with  $D_F(t-t')=m\langle 0|\hat{T}[q_I(t)q_I(t')] |0\rangle$ the \emph{Feynman propagator} (one can regard this system as a $0$-dimensional field theory, in which case one usually set $m=1$, with the role of mass played by $\omega$). Here $|0\rangle$ is the ``free'' vacuum, i.e. the ground state of the quadratic part of the Hamiltonian and $q_I(t)$ is a conventional position operator evolved without the source. Note that, in analogy with Eq.\ (\ref{eq:thcorrel}), there is no need to specify the state $|0\rangle\langle 0|$ in the definition of $Z_\tau[j]$. 
   
    The functional derivatives of $Z_\tau[j]/Z_\tau[0]$ appear now linked to a variation of the operator $e^{i\mathcal{S}_\tau[j]}$ providing general $\tau$-invariant expressions for \emph{vacuum correlation functions}. In particular,
    \begin{equation}\label{eq:propf}
     D_F(t_1-t_2)= m\,\hat{T}_\tau\int_{0}^{\tau} \frac{d\tau_1}{\sqrt{\tau}}\int_{0}^{\tau} \frac{d\tau_2}{\sqrt{\tau}}
      \frac{{\rm Tr}[e^{i\mathcal{S}_\tau[0]} q_I(t_1,\tau_1)q_I(t_2,\tau_2)]}{{\rm Tr}[e^{i\mathcal{S}_\tau[0]}]}\,.
    \end{equation}
    To obtain Eq.\ (\ref{eq:propf}) one can write
\begin{equation}\label{eq:tauevolution}
\begin{split}
       e^{i\mathcal{S}_\tau[j]}
       =e^{i\mathcal{S}_\tau[0]}\,\hat{T}_{\tau}\exp\Big(i\medop{\int dt \int_{0}^{\tau} \frac{d\tau'}{\sqrt{\tau}}}\sqrt{m}j(t)q_I(t,\tau')\Big)
\end{split}
\end{equation}
    for $q_I(t,\tilde{\tau}):=e^{-i\mathcal{S}_{\tau}[0]}q(t)e^{i\mathcal{S}_\tau[0]}$ and derive with respect to $j(t)$ at the operator level (the time ordering in (\ref{eq:propf})-(\ref{eq:tauevolution}) is applied to the parameter $\tau$). 
    In general, this procedure shows that for each operator on the correlation function we must insert an operator in the proper Hilbert, in analogy with Fig.\ (\ref{fig1}). Then we integrate over each $\tau_i$ preserving the $\tau$-ordering.
   
For small $\tau$ the form of Eq.\  (\ref{eq:correl}) is recovered from (\ref{eq:propf}), with operators $q_I(t,\tau)\to q(t)$  inserted at the evolution time:  
\begin{equation}\label{eq:propintf}
    \frac{\int \mathcal{D}q\,e^{iS_{\rm cl}}q(t_1)q(t_2)}{\int \mathcal{D}q\,e^{iS_{\rm cl}}}=\lim_{\tau\to 0}\frac{{\rm Tr}[e^{i\mathcal{S}_\tau[0]} \sqrt{\tau}q(t_1)\sqrt{\tau}q(t_2)]}{{\rm Tr}[e^{i\mathcal{S}_\tau[0]}]}\,.
\end{equation}
Moreover,  the description of general theories in this limit corresponds to trace expressions 
with general actions $\mathcal{S}_\tau$ (Eq.\ (\ref{eq:tauaction})), in close analogy with the PI definitions and in agreement with the discussion of section \ref{sec:consmalltau}. In particular, the replacement of an interacting QA in (\ref{eq:propintf}), yields the corresponding  interacting propagator and associated Feynman rules.

\subsection{Spacetime states and large  $\tau$ limit \label{IIIC}}
\subsubsection{Extended trace expressions as  spacetime vacuum  mean values}
The possibility of a useful definition of states in spacetime\footnote{To be read as ``states in spacetime'', e.g. states representing field configurations in spacetime (in contrast with configurations in space); not to be confused with states of spacetime itself, which is not quantized in this work.} scenarios 
has been recently explored in the literature \cite{ish.94,cot.18,fit.15, ho.17}. This has prompted discussions on potential modifications, either in the axioms that define a state \cite{fit.15, ho.17} or in the nature of the  Hilbert space considered \cite{ish.94,cot.18}.
The $\tau$-invariance property allows us to consider a new possibility: in 
the $\tau\gg 1$ limit we have
\begin{equation}\label{eq:vacuumproj}
    e^{i\mathcal{S}_\tau[0]}\to |\Omega\rangle \langle \Omega|\,,
\end{equation}
i.e. the exponential of the spacetime quantum action becomes a projector onto the \emph{spacetime vacuum} of the free theory, in analogy with the $\T \to \infty$ limit of a conventional time evolution operator $e^{-iHT}\to |0\rangle \langle 0|$, with $|0\rangle$ the free \emph{non-extended} vacuum of $H$ (however the $\tau$-limit does not require $\T\to \infty$). Then, for $\tau$-invariant quantities the associated sums over histories, which in principle involve a complete basis of spacetime states, can be reduced to single spacetime vacuum expectation values. The $\tau$-invariance is thus revealing a ``continuum interpolation'' between these two apparently different notions. 

In particular, for the generating functional of the previous section, Eqs.\  (\ref{eq:genfeyn}), (\ref{eq:tauevolution}) and (\ref{eq:vacuumproj}) imply

\begin{align}\label{eq:generatingvac}
    \frac{Z_\tau[j]}{Z_\tau[0]}&=\frac{1}{Z[0]}
    \int \mathcal{D}q\, e^{i S_{\rm cl}[j]}
    \\
    &=\lim_{\tau\to \infty} \langle \Omega|\hat{T}_{\tau}\exp\Big(i\!\medop{\int\! dt\! \int_{0}^{\tau} \frac{d\tau'}{\sqrt{\tau}}}\sqrt{m}j(t)q_I(t,\tau')\Big) |\Omega\rangle\,.
\end{align}
We see that the ``normalized'' generating functional is a pure spacetime vacuum mean value (with $Z_\tau[0]\to 1$).   

Similar considerations hold for the Feynman propagator, and for any other quantity related to generating functionals, with Eqs.\ (\ref{eq:propf}) and (\ref{eq:vacuumproj}) implying
   \begin{equation}\label{eq:proplim}
   \begin{split}
        \langle 0|\hat{T}q_I(t_1)q_I(t_2)|0\rangle &=\lim_{\tau\to \infty}\;\hat{T}_\tau\int_{0}^{\tau} \frac{d\tau_1}{\sqrt{\tau}}\int_{0}^{\tau} \frac{d\tau_2}{\sqrt{\tau}}\,
        \langle \Omega| q_I(t_1,\tau_1)q_I(t_2,\tau_2) |\Omega\rangle\,.
   \end{split}
   \end{equation}
Notice that in the l.h.s.\ the interaction picture in $\mathfrak{H}$ corresponds to the evolution in $t$ while the interaction picture in $\mathcal{H}$ to the evolution in $\tau$ (while $t$ indicates the Hilbert space of $q(t)$). 
It is also easy to see that $\theta(t_1\!\shortminus t_2)\langle 0|q_I(t_1)q_I(t_2)|0\rangle =\lim\limits_{\tau\to \infty} \int_0^\tau \frac{d\tau_1}{\sqrt{\tau}}\int_0^\tau\frac{d\tau_2}{\sqrt{\tau}}
\theta(\tau_1\!\shortminus\tau_2)\langle \Omega|q_I(t_1,\tau_1)q_I(t_2,\tau_2)|\Omega\rangle$.
The propagator for other theories can be defined from these basic elements and related to vacuum mean values as well.

\subsubsection{Extended states in relativistic quantum field theories}
We can improve the relation (\ref{eq:proplim}) by removing the explicit 
limit on $\tau$ and leaving only one integral on the variable $\tau_1-\tau_2$. 
Remarkably, the result relates sp states in $\mathcal{H}$ to those  considered in string theory inspired approaches (and other quantum time formalisms, as suggested in \cite{diaz.21}). For a proper comparison it is appropriate to work in $D=d+1$ spacetime dimensions by simply replacing $q(t)\to\phi(x)$, $p(t)\to \pi(x)$ such that the algebra (\ref{eq:alg}) yields 
\begin{equation}\label{eq:algfield}
    [\phi(x),\pi(y)]=i\delta^{(d+1)}(x-y)\,.
\end{equation}
Notice that (\ref{eq:algfield}) is \emph{not} an equal time commutation relation as the conventional one in $d$ spatial dimensions: the ``extra'' delta corresponds to the time dimension (see Eq.\ (\ref{eq:algcont})).  This should not be confused with the canonical quantization of a classical theory with an extra dimension: it is not the number of dimensions which is modified, but the Hilbert space construction and the ensuing quantization scheme. We can nonetheless speculate that the $\tau$ parameter might be treated as a ``holographic'' coordinate connecting the $d+1$ theory with a $d+2$ canonical theory (a little analysis shows that the extra-dimensional theory must be highly non-local in order to handle interactions).

We also take the opportunity to briefly discuss Lorentz invariance at the Hilbert space level: the new algebra is explicitly covariant for $$U^\dag(\Lambda)\phi(x)U(\Lambda)=\phi(\Lambda x)\,, \;\;\;\;U^\dag(\Lambda)\pi(x)U(\Lambda)=\pi(\Lambda x)\,.$$
 with $U(\Lambda)$ the unitary sp transformation 
 associated to the Lorentz transformation $\Lambda$ \cite{dia.19}. Therefore, 
Lorentz transformations are defined geometrically in $\mathcal{H}$, in analogy with rotations and independently of the dynamics. 
Moreover, if we introduce a ``second order''  QA 
\begin{equation}\label{eq:action2}
\begin{split}
    \mathcal{S}^{(2)}_{\tau}[j]&:=-\tau\int d^{D}x \,\big[ A^\dag(x) (\partial^2+m^2)A(x)-
    \sqrt{\tfrac{m}{\tau}}j(x)\phi(x)\big]\,,
\end{split}
\end{equation}
with $A(x):=(\sqrt{m}\phi(x)+i\pi(x)/\sqrt{m})/\sqrt{2}$,  it is clear that $[\mathcal{S}^{(2)}_{\tau}[0],U(\Lambda)]=0$ (see also \cite{diaz.21}). The local operators $A(x)$ satisfy   $[A(x),A^\dag(y)]=\delta^{(d+1)}(x-y)$.  This brief introduction of Lorentz covariance shows that the spacetime canonical formulation within $\mathcal{H}$ allows one to preserve spacetime symmetries explicitly, an  advantage previously exclusive to the PI formulation. It also shows that more general forms of  QAs can be introduced \footnote{While the $d+1$ generalization of $S_\tau[j]$ and related results is straightforward, we are employing the second order action to preserve Lorentz covariance explicitly at all steps. This can be achieved with $S_\tau[j]$ also but it requires a proper discussion of the Legendre transformation's time choice, see \cite{DMR.24}.}.

With these conventions the field can be expanded as
\begin{equation}\label{eq:field}
    \phi(x)=
    \int \frac{d^{D}p}{\sqrt{(2\pi)^{D}2m}}\left(e^{ipx}A(p)+e^{-ipx}A^\dag(p)\right)\,,
\end{equation}
with the operators $A(p)$ the FT of $A(x)$, which are those bringing ${\cal S}^{(2)}_\tau[0]$ 
to its normal form:
\begin{equation}\label{eq:action22}
\begin{split}
    \mathcal{S}^{(2)}_{\tau}[j]&:=\tau\int d^{D}p\,\big[(p^2-m^2) A^\dag(p)A(p) 
    +\tfrac{j(-p)}{\sqrt{2\tau}}A(p)+\tfrac{j(p)}{\sqrt{2\tau}}A^\dagger(p)\big]\,.
\end{split}
\end{equation}
All previous results related to the generating functional, including the $\tau$-invariance hold in complete analogy:
defining as before   $Z^{(2)}_\tau[j]={\rm Tr}[e^{i\mathcal{S}^{(2)}_{\tau}[j]}]$,  we obtain 
\begin{equation}
   Z^{(2)}_\tau[j] =Z^{(2)}_\tau[0]\exp\Big(-\tfrac{1}{2}\medop{\int} d^{D}x d^{D}y\, j(x)D_F(x-y)j(y) \Big)\,,
\end{equation}
a multidimensional generalization of Eq.\ (\ref{eq:scl}), with the Feynman propagator $D_F(x-y)=\langle 0|\hat{T}\phi_I(x)\phi_I(y)|0\rangle$ appearing now explicitly (as usual we are setting $m^2\to m^2-i\epsilon$).   
Moreover, a new version of Eq.\ (\ref{eq:vacuumproj}) is obtained:
\begin{equation} e^{i\mathcal{S}^{(2)}_\tau[0]}\to |\Omega\rangle \langle \Omega|\label{eq:spv2}
\end{equation}
when $\tau \to \infty$ and where $|\Omega\rangle$ is the spacetime of $A(x)$ satisfying $A(x)|\Omega\rangle=A(p)|\Omega\rangle=0$, $U(\Lambda)|\Omega\rangle=|\Omega\rangle$. By separating the parts of the QA with and without the source, as in Eq.\ (\ref{eq:tauevolution}), we obtain 
\begin{equation}\label{eq:proplim2}
   \begin{split}
        \langle 0|\hat{T}\phi_I(x)\phi_I(y)|0\rangle &=\lim_{\tau\to \infty}\;\hat{T}_\tau\int_{0}^{\tau} \frac{d\tau_1}{\sqrt{\tau}}\int_{0}^{\tau} \frac{d\tau_2}{\sqrt{\tau}}
        \,m\,
        \langle \Omega| \phi_I(x,\tau_1)\phi_I(y,\tau_2) |\Omega\rangle\,,
   \end{split}
 \end{equation}
for $\phi_I(x,\tau):=e^{-i\mathcal{S}_{\tau}^{(2)}[0]}\phi(x)e^{i\mathcal{S}_{\tau}^{(2)}[0]}$, the field operator ``evolved'' with the second order action. It is straightforward to show that both orderings of $\tau$ yield half the propagator ($D_F(x-y)=D_F(y-x)$), while  Lorentz invariance is manifest on the r.h.s.\ since $\phi_I(\Lambda x,\tau)=U^\dag(\Lambda)\phi_I(x,\tau)U(\Lambda)$ and the spacetime vacuum is invariant. 

The result (\ref{eq:proplim2})
is the $D$-dimensional version of  (\ref{eq:proplim}) which can be compared with string theory-like expressions: by using Eq.\ (\ref{eq:field}) the integrand in  (\ref{eq:proplim2}) can be written for $\tau_1>\tau_2$ as
$      \langle \Omega| A(x)e^{i(\tau_1-\tau_2)\mathcal{J}^{(2)}}A^\dag(y)|\Omega\rangle/2\,,
$
with \begin{equation}
\mathcal{J}^{(2)}:=\tau^{-1}\mathcal{S}^{(2)}_\tau[0]=\int d^{D}p\,(p^2-m^2) A^\dag(p)A(p) \,,\label{eq:j2}
\end{equation}
independent of $\tau$. Then, since the integrand in (\ref{eq:proplim2})
 depends only on the difference $\tau_1-\tau_2$ we find 
    \begin{align}\label{eq:propstr}
        \langle 0|\hat{T}\phi_I(x)\phi_I(y)|0\rangle 
        &=\int_{0}^\infty d\tau\, \langle \Omega| A(x)e^{i\tau\mathcal{J}^{(2)}}A^\dag(y) |\Omega\rangle\nonumber\\
        &=\int_{0}^\infty d\tau\, \langle x|e^{i\tau\mathcal{J}^{(2)}}|y\rangle
    \end{align}
where we have defined the single particle (sp) states 
\begin{equation}
|x\rangle:=\sqrt{2m}\phi(x)|\Omega\rangle=A^\dag(x)|\Omega\rangle\,.
\end{equation}
Remarkably, the extended Fock space result (\ref{eq:propstr}) involves the operator $\mathcal{J}^{(2)}$, which is the ``second quantizatized'' version of $\mathcal{J}^{(2)}_{\text{sp}}=P^2-m^2$  (see Eq.\ (\ref{eq:j2}) and \cite{dia.19,diaz.21}) defining the mass-shell condition of parameterized particles \cite{pam.50,Ma.00,ch.22} through   $\mathcal{J}^{(2)}_{\text{sp}}|\Psi\rangle=0$ on 
$\mathfrak{H}=L^2(\mathbb{R}^{d+1})$.
Thus, for two-point contractions the form \eqref{eq:propstr}  
 reduces to the well-known ``worldline'' expression of the propagator \cite{Ma.97,Schb.01} $\langle 0|\hat{T}\phi_I(x)\phi_I(y)|0\rangle=\int_{0}^\infty d\tau\, \langle x|e^{i\tau(P^2-m^2+i\epsilon)}|y\rangle$, which involves just sp states (in first quantization).  
 
 We point out that in the current approach 
previous results emerge from a fully-fledged spacetime formulation of PIs and quantum (scalar) field theory correlation functions.  The \emph{excitations of the fields are} now \emph{spacetime states}.  Whereas further development exceed the scope of this manuscript, all basic ingredients for developing general (interacting) theories are already contained in it: on the one hand, one can introduce (as usual) interacting theories through functional variations of the generating functional. At the extended Hilbert space level this defines new $\tau$-invariant generalizations of interacting QAs, reducing to conventional diagonal in time actions only for small $\tau$ (besides the case of a  ``linear interaction'' considered in previous sections).  
On the other hand, physical quantities arise from correlation functions essentially through full $d+1$ FTs (e.g., the LSZ reduction formula  \cite{srednicki2007}). While in the conventional formulation the FT in time is related to unitary evolution, here such FTs naturally lead to the nonlocal in time operators $A^\dag(p)$ which diagonalize $\mathcal{S}_\tau^{(2)}[0]$. 

Moreover, the momenta of each external particle involved in S-matrix elements satisfy the on-shell condition. This leads to extended creation (annihilation) operators which are ``stationary'' in $\tau$ ``evolution'', i.e. \begin{equation}[\mathcal{S}^{(2)}_\tau[0],A^\dag(E_{pm},\textbf{p})]=0\,,\end{equation} precisely the condition introduced in \cite{dia.19,diaz.21} 
on operators creating (free) \emph{physical states} by acting on $|\Omega\rangle$. Such states arise then naturally for large $\tau$ in the extended formulation of scattering theories and represent the external (asymptotically free) particles. 

It is worth mentioning that a similar constraint (but not written in terms of QAs) has been recently introduced in \cite{g.22} for non-interacting theories in the context of a formulation of relativistic QM in terms of events.  Besides some fundamental differences in interpretation \cite{g.22} (we note however that a trajectory can be considered as a set of events, hence the Hilbert space is the same), our present results provide a clear route to introduce interactions 
in this new related formulation as well. Moreover, on the basis of these results, and in the scenario of small $\tau$ where the interacting actions are local (see above), it has been recently shown how to treat interacting field theories in this extended setting \cite{DMR.24}. A definition of spacetime state, valid for small $\tau$ and for discrete time with proper modifications has also been proposed therein.

  \section{Conclusions\label{sec:IV}}
  We have provided a full quantum formulation of  Feynman PIs on the basis of an  extended spacetime Hilbert space and a concomitant QA. Fundamental expressions can be cast as spacetime traces, and different PI formulations emerge  naturally from the use of different extended bases. Standard representations correspond to  trajectory-like product-in-time bases (e.g. coordinate and coherent trajectory-states), but the formalism makes non-local in-time bases also accessible. In particular, Fourier and  Mastubara-like evaluations are special instances of the latter, naturally arising here through the eigenbasis of quadratic QA operators.  
  We have also discussed the general diagonalization of the QA for interacting  theories  (Appendix C), 
  showcasing the entangled-in-time nature of its eigenstates, which lie beyond those of a Fourier (in time) Fock basis.

  In the continuum time case (section \ref{sec:continuum}) this allows one to define and manipulate trace expressions without the subtleties of conventional PIs, while the connection with classical physics may still be discussed within the operator framework (see  \ref{ApD}). The possibility of new regularization schemes also arises. 
  Moreover, a timescale invariance becomes now apparent in the new expressions for correlation functions, which leads to a direct connection between a given QA and the corresponding spacetime vacuum, as  shown in  \ref{IIIC}. When applied to quantum fields,  expressions from 
  the first quantization string-inspired approach \cite{Schb.01} and/or the relativistic PaW formalism described in \cite{di.19,dia.19}
  are recovered 
  at the one particle level. 
  
  From a wider perspective, the present results constitute an important step in the development of general spacetime symmetric extensions of QM: through the new representation of PIs, a spacetime symmetric Hilbert space representation of any conventional theory is formally achieved, including the case of interacting quantum field theories (see considerations in section \ref{IIIC}).
 A new route for a proper definition of physical spacetime states was also unveiled by exploiting the aforementioned large $\tau$ limit. Interestingly, even for finite or small $\tau$, and in the discrete time scenario, a notion of state can be assigned to the previous representation \cite{DMR.24}: the essential idea is to treat the exponential of the action as a thermal-like state (see some of the remarks in section \ref{sec:corrfunc}). This can be developed through a ``generalized'' purification technique recently introduced in the context of holographic dualities with the aim of discussing time-like entanglement \cite{har.22,nar.22} (in conventional, nonextended QM).

In this same scenario, additional novel 
 possibilities emerge, such as the consideration of nonseparable-in-time interactions, the emergence of quantum time operators and energy-time uncertainty relations, and the rigorous definition of entanglement in time: in the same way as standard second quantization is required for the notion of a reduced density matrix of a space interval, and hence for entanglement in space \cite{muss.22}, the present second quantized spacetime states formalism is a natural scenario to accommodate the notion of entanglement in time. 
 At the same time, the present formalism opens the door to novel computational techniques for PI evaluation. In particular, the conventional ``sum over histories'', previously only accessible through classical computations, now admits the application of quantum protocols for trace evaluation (\ref{ApAA}).  These aspects are currently under investigation. 

\vspace{0.8cm}
	\section*{Acknowledgements}
 The authors would like to thank Marco Cerezo and Diego García-Martín for fruitful discussions. N.L.D was supported by the Laboratory Directed Research and Development (LDRD) program of Los Alamos National Laboratory (LANL) under project numbers 20230049DR, 20230527ECR, and by the Center of Nonlinear Studies under project number 20250614CFR-NLS. 
		We also acknowledge support from CONICET (N.L.D., J.M.M.) and  CIC (R.R.) of Argentina.  Work supported  by CONICET PIP Grant 11220200101877CO.

\appendix
\section{General $T$ in propagators\label{ApA}}
In the discrete construction developed in the main text,
we have considered $\N=\T/\epsilon$ copies of the original Hilbert space while identifying $\T$ with the amount of evolution of final states. Here we discuss the more general situation which arises from relaxing this identification in the case of the bosonic particle.

We note first that Eq.\ \eqref{eq:PI}  of the main body holds also in $\mathcal{H}'=\otimes_{t=0}^{N'-1}\mathfrak{H}_t$ with $\N'>\N$ and $\mathcal{S}$ still defined as in Eq.\ \eqref{eq:S} but with $\mathcal{P}_t$ the
generator of time translations in $\mathcal{H}'$. This follows from replacing in the right-hand side of Eq.\ \eqref{eq:tslices}  $\N\to \N'$,
$$\otimes_{t=0}^{N-1}e^{-iH\epsilon}\equiv \otimes_{t=0}^{N-1}e^{-iH\epsilon}\otimes_{t>N-1}\mathbbm{1}_t$$ and integrating over the variables $q_{t>N-1}$ such that the equality holds. Then in Eq.\ \eqref{eq:PI}  $\N\to \N'$ but not in the product of Eq.\ \eqref{eq:S}. 
This invariance allows one to discuss any time evolution of interval $\T<\T'=\N'\epsilon$ (with any origin)  within a single extended space $\mathcal{H}'$. 

In particular, by considering $\mathcal{H}'=\mathcal{H}\otimes \mathfrak{H}_{N}$ 
\begin{equation}\label{eq:schr}
    \langle q',\T+\epsilon|q\rangle-\langle q',\T|q\rangle={\rm Tr}_{\mathcal{H}'}\left[[e^{i\mathcal{S}'}-e^{i\mathcal{S}}]|q\rangle_0\langle q'|\right] 
\end{equation}
where 
$e^{i\mathcal{S}'}=e^{i\mathcal{S}}e^{-i\epsilon H_N}=e^{-i\epsilon H_0}e^{i\mathcal{S}}$. By writing then
\begin{equation}
    e^{i\mathcal{S}'}-e^{i\mathcal{S}}=[e^{-iH_0\epsilon}-\mathbbm{1}]e^{i\mathcal{S}}\,,
\end{equation}
and applying Eq.\ \eqref{eq:PI}  of the main body to the right-hand side of Eq.\ (\ref{eq:schr}),
 the ``discrete'' \emph{Schrodinger equation} 
 \begin{equation}
     \langle q',\T+\epsilon|q\rangle-\langle q',\T|q\rangle=\langle q',\T|[e^{-iH\epsilon}-\mathbbm{1}]|q\rangle
 \end{equation}
  is recovered. The continuum limit follows of course by dividing both members by $-i\epsilon$ such that for $\epsilon\to 0$ the left-hand side is $i$ times the time derivative of $\langle q',\T|q\rangle $ while 
$i[e^{-iH\epsilon}-\mathbbm{1}]/\epsilon\to H$.

  Note also that Eq.\ (\ref{eq:schr}) has exactly the form of \emph{Schwinger's actions principle} \cite{sch.51} which relates general variations of $\langle q',\T|q\rangle$ to the matrix elements of  variations of the Schwinger's action operator. However, 
  in Schwinger's formulation a complete set of commuting operators is available on space-like surfaces (at a given time). From the canonical point of view, its QA involves operators in the Heisenberg picture for which no extended algebra apply \cite{diaz.21}, a fundamental difference with the present construction.

\section{Proof of the relation between $\mathcal{S}$, $\mathcal{P}_t$ and $\mathcal{V}$, and correlation functions \label{ApB}}
In this section we will prove equation \eqref{eq:unitarytrans} of the main body. 
We will first establish the equivalence between that result and the following expression:
\begin{equation}\label{c1}
    e^{-i\mathcal{P}_t\epsilon}\mathcal{V}^\dag e^{i\mathcal{P}_t\epsilon}=U^\dag_{N-1}(\T) \otimes_t U_t[(t+1)\epsilon,t\epsilon]\mathcal{V}^\dag 
\end{equation}
\begin{proof}
The proof of the equivalence follows immediately by rewriting (\ref{c1}) as $$\mathcal{V}^\dag e^{i\mathcal{P}_t\epsilon}\mathcal{V}=e^{i\mathcal{P}_t\epsilon}U_{\N-1}^\dag(\T)\otimes_t U_t[(t+1)\epsilon,t\epsilon]\,.$$ Now, considering that $e^{i\mathcal{P}_t\epsilon}U_{\N-1}^\dag(\T)e^{-i\mathcal{P}_t\epsilon}=U^\dag_0(\T)$ we can write $$U_0(\T)\mathcal{V}^\dag e^{i\mathcal{P}_t\epsilon}\mathcal{V}=e^{i\mathcal{P}_t\epsilon}\otimes_t U_t[(t+1)\epsilon,t\epsilon]=e^{i\mathcal{S}}\,,$$ which is precisely Eq.\ \eqref{eq:unitarytrans} of the main body. Note that we are considering the general definition of $e^{i\mathcal{S}}$ (possibly time-dependent).
\end{proof}

The proof of Eq.\ \eqref{eq:unitarytrans} in the main body now reduces to proving (\ref{c1}).
\begin{proof}
The action of the translation operator on $\mathcal{V}^\dag=\otimes_{t=0}^{\N-1}U_t(t\epsilon)$ in the left-hand side of (\ref{c1}) yields
\begin{align}\label{c2}
   e^{-i\mathcal{P}_t\epsilon}\otimes_{t=0}^{\n-1}U_t(t\epsilon) e^{i\mathcal{P}_t\epsilon}&=\otimes_{t=0}^{\n-1}U_{t-1}(t\epsilon)\nonumber=\otimes_{t=0}^{\n-2}U_t[(t+1)\epsilon]\nonumber\\&=U_{\n-1}^{\dag}(\T)\otimes_{t=0}^{\n-1}U_t[(t+1)\epsilon] 
\end{align}
where we used $U(0)=\mathbbm{1}$ and $\T=\N\epsilon$. On the other hand,
\begin{align}\label{c3}
    \otimes_{t=0}^{\n-1}U_t[(t+1)\epsilon]\mathcal{V}&=\otimes_{t=0}^{\n-1}U_t[(t+1)\epsilon]\otimes_{t=0}^{\n-1}U^\dag_t(t\epsilon)\nonumber\\&=\otimes_{t=0}^{\n-1}U_t[(t+1)\epsilon]U^\dag_t(t\epsilon)=\otimes_{t=0}^{\n-1}U_t[(t+1)\epsilon,t\epsilon]\,.
\end{align}
By multiplying (\ref{c2}) on the right by $\mathcal{V}\mathcal{V}^\dag=\mathbbm{1}$ and using (\ref{c3}) we recover (\ref{c1}).
\end{proof}

We describe now how the previous result allows for a straightforward derivation of the correlation functions expressions such as \eqref{eq:correl}. We recall that  one can map conventional traces of composition of operators into spacetime traces of tensor product of operators by adding the time translation operator $e^{i \mathcal{P}_t\epsilon}$, as shown in \eqref{eq:matrixel}. On the other hand, it is clear that conjugating tensor products of operators with $\mathcal{V}$ corresponds to evolving them (see \eqref{eq:ophe}). Putting all these results together we can write
\begin{equation}
\begin{split}&
\langle i',\T|O^{(N-1)}_H(\T)\ldots O^{(2)}_H(2\epsilon)O^{(1)}_H(\epsilon)O^{(0)}_H(0)|i\rangle=\\
    &= {\rm Tr}_{\cal H}\big[e^{i \mathcal{P}_t\epsilon}\otimes_t O^{(t)}_H(t) |i\rangle_0\langle  i',T|\big]\\
    &={\rm Tr}_{\cal H}\big[e^{i \mathcal{P}_t\epsilon}\mathcal{V}(\otimes_t O^{(t)}) \mathcal{V}^\dag|i\rangle_0\langle  i'|U_0(\T)\big] \\
     &={\rm Tr}_{\cal H}\big[\underbrace{U_0(\T)\mathcal{V}^\dag e^{i \mathcal{P}_t\epsilon}\mathcal{V}}_{e^{i\mathcal{S}}}\otimes_t O^{(t)} |i\rangle_0\langle  i'|\big]\,,
\end{split}
\end{equation}
where in the last equality we recognize the combination of $\mathcal{V}$ and $e^{i\mathcal{P}_t \epsilon}$ which gives rise to $e^{i\mathcal{S}}$ according to the previous theorem. Notice that one can set operators to be equal to the identity, so that only some of them actually appear in the l.h.s.\ which on the r.h.s.\ corresponds to particular ``insertions''. Let us also remark that this result is general, and can be applied to any system. In particular, it can be applied to general bosonic systems, such as fields, and hence the multidimensional case discussed in section \ref{sec:corrfunc}, which exhibits spacetime symmetry is also included.

\section{Structure of the QA operator for general \\
time-independent Hamiltonians \label{Apkerr}}
Here we discuss the eigenstates and eigenvalues of the quantum action operator for a general time-independent theories,  as a function of the spectrum of the Hamiltonian. The purpose is to discuss the general structure of the QAs and how non-quadratic actions lead to states with non-classical features.

For $H|n\rangle=\lambda_{n}|n\rangle$ with $\{|n\rangle\}$ the eigenbasis of $H$, 
we can define the product-in-time basis $|n_0,n_1,\dots\rangle$  satisfying for each time slice the eigenvalue equation $H_t |n_0,n_1,\dots\rangle=\lambda_{n_t}|n_0,n_1,\dots\rangle$. With this auxiliary definition, we now 
introduce the orthonormal basis of states given by 
\begin{equation}
    |k;n_0n_1,\dots, n_{N-1}\rangle:=\frac{1}{\sqrt{M}}\sum_{l=0}^{M-1}e^{-i\epsilon \frac{N}{M}\omega_k l }e^{i l \epsilon\mathcal{P}_t}\,|n_0,n_1,\dots, n_{N-1}\rangle\,,
    \label{C1}
\end{equation}
with
  $M$ the lowest integer such that  
the product state  remains invariant 
  under application of an $M$-step time translation   $e^{iM\epsilon {\cal P}_t}$ ($1\leq M\leq N$, with $N=ML$ and $M,L$ integer)
  and  $k=0,\ldots,M-1$. For instance, a state $|n,n,\ldots,n\rangle$ corresponds to $M=1$ (and hence $k=0$), $|n_0,n_1,n_0,n_1\ldots\rangle$ to $M=2$ (assuming $N$ even), etc. 
Since $[{\cal P}_t,\sum_t H_t]=0$, it is straightforward to show that 
\begin{equation}    e^{i\mathcal{S}}|k;n_0,n_1,\dots, n_{N-1}\rangle=\left( e^{i\epsilon\frac{M}{N}\omega_k}e^{-i\sum_t \epsilon \lambda_{n_t}}\right) |k;n_0,n_1,\dots, n_{N-1}\rangle
\end{equation}
for any QA of the form $\mathcal{S}=\epsilon \mathcal{P}_t-\epsilon \sum_t H_t$. This result applies for bosons as well as for general finite dimensional systems. Conceptually, we see that we can always think of each of these states as a sum over all possible translations in time of a basic product-in-time history with the addition of all possible phases for each step, thus carrying the ``sum over histories'' of the PI formulation to the eigenstate structure.  Eqs.\ \eqref{eq:correl} and \eqref{eq:matrixel} can thus be verified for  a general constant $H$. 
One can explicitly see that for general mean values all of the states \eqref{C1} are relevant, as one can think of the set of states \eqref{C1} as covering all possible ``trajectories in energy space''.

Let us now consider the case of conventional bosons. Interestingly, this eigenbasis is generally neither ``$t$'' nor ``$n$'' local (the Fourier basis), at least for general states involving more than one boson. This can be seen already by considering the basic example of two time slices and the Kerr Hamiltonian $H_K$ (see the QA in \eqref{eq:actionkerr} and discussion below). In fact, consider the eigenstates
 $|k;n_0n_1\rangle=\frac{|n_0n_1\rangle+(-1)^k |n_1n_0\rangle}{\sqrt{2}}$ for $|n_0n_1\rangle=\frac{A_{0}^{\dag n_0}}{\sqrt{n_0!}}\frac{A_{1}^{\dag n_1}}{\sqrt{n_1!}}|\Omega\rangle$ and $n_0\neq n_1$.  For $n_0=0$ and $n_1>2$, the states $|k;n_0n_1\rangle$ are  ``entangled'' Noon-like states \cite{Noon1,Noon2} which cannot be always written as permanents in any sp basis. 
In order to show this feature, let us introduce the single particle reduced density matrix $R$ whose entries are defined as follows $R_{tt'}:=\langle \psi| A_t^\dag A_{t'}|\psi\rangle$ (i.e. the standard definition of a reduced density matrix, but with temporal modes). For the example $|\psi\rangle=|k;03\rangle$ one obtains \begin{equation}
    R=\begin{pmatrix}
        3/2 &0\\
        0& 3/2
    \end{pmatrix}\,.
\end{equation}
Since $R$ is already diagonal, and under a unitary transformation of the modes the matrix $R$ transforms linearly, there cannot be a sp basis in which e.g. $|\psi\rangle\sim B_0^\dag B_1^{\dag 2}|\Omega\rangle$ for some creation operators $B_i$ (otherwise $R$ would have integer eigenvalues; for example $B_0^\dag B_1^{\dag 2}|\Omega\rangle$ leads to $R=\text{diag}(1,2)$, while $B_0^{\dag 2}|\Omega\rangle$ to $R=\text{diag}(3,0)$). These considerations clearly hold for general states with arbitrary $N$ (beyond the NOON state case), with e.g. $$|0;2,0,0\ldots\rangle=\frac{(A_0^{\dag 2}+A_1^{\dag 2}+A_2^{\dag 2}+\ldots)}{\sqrt{N}}|\Omega\rangle$$ leading to $R=\text{diag}(2/N,2/N,2/N,\dots )$. In summary, we have shown that, even if $H_K$ has an eigenbasis of Fock states, the QA $\mathcal{S}_K$ does not (unless some additional degeneracy exists, typical of non-interacting theories; see below). This notable result is a consequence of having diagonalized both an interacting Hamiltonian  and  the time translation operator simultaneously.

For completeness, let us also explain how the previous complicated structure simplifies in the case of quadratic Hamiltonians, thus allowing Matsubara like expansions and the existence of a Fock eigenbasis: as the eigenvalues are linear in $n$, i.e. $\lambda_n=\omega n $ one finds a large degeneracy. For example, $|0;12\rangle$ and $|0;03\rangle$ lead to the same eigenvalue $e^{-i3\omega \epsilon}$ of $e^{i\mathcal{S}}$. One can exploit this degeneracy to write all the states of $3$ bosons in the Fourier basis as linear combinations within this subspace, so that they are eigenstates as well. Instead for $H_K$ the corresponding eigenvalues are $e^{-i(3\omega  \epsilon+5\lambda  )}$ and $e^{-i(3\omega  \epsilon+9\lambda  )}$ so that  the previous ``accidental'' degeneracy is broken.

Let us finally remark that under the approximation $\epsilon\ll 1$ one can generalize the previous considerations to time-dependent Hamiltonians by choosing  the initial product-in-time basis such that 
at each time step the eigenstate of the Hamiltonian at that fixed time is considered.

\section{About the ``partition function'' of the harmonic oscillator in the continuum limit \label{ApC}}

We discuss here the trace of $e^{i\mathcal{S}_\tau}$ for  continuum time and for
\begin{align}
    \mathcal{S}_\tau&= \tau\sum_{n=-\infty}^{\infty} (\omega_n-\omega+i\tilde{\epsilon} \omega_n^2)A^\dag_n A_n \label{d1}\\
    &= \tau\int dt\,A^\dag(t)(i\partial_t-\omega-i\tilde{\epsilon} \partial^2_t)A(t)\label{d2}\,.
\end{align}
Note that we have introduced a convergence factor $\propto\tilde{\epsilon}$ (for the moment $\tilde{\epsilon}\in \mathbb{R}$).

We can immediately compute the trace in the Fourier Fock basis ($\tilde{n}\in \mathbb{N}$ is the number of occupation of a certain mode $n$) obtaining
\begin{align}\label{eq:geomseries}
     {\rm Tr}[e^{i\mathcal{S}_\tau}]&=\sum_{\tilde{\textbf{n}}}\langle \tilde{\textbf{n}}|e^{i\mathcal{S}_\tau}|\tilde{\textbf{n}}\rangle=\medop{\prod}_{n=-\infty}^{\infty}\,\medop{\sum}_{\tilde{n}_n=0}^\infty e^{i\tau (\omega_n-\omega+i\tilde{\epsilon} \omega_n^2)\tilde{n}_n}\nonumber\\
     &=\medop{\prod}_{n=-\infty}^{\infty}\Big[1-e^{i\tau (\omega_n-\omega+i\tilde{\epsilon} \omega_n^2)}\Big]^{-1}\,, 
\end{align}
where in the last step we assumed $|e^{i\tau [\omega_n-\omega+i\tilde{\epsilon} \omega_n^2]}|<1 \, \forall n$. This is strictly true for the modes $n\neq 0$. For the $0$-mode this holds for $\omega$ slightly imaginary (as usual), while for $\omega \in \mathbb{R}$ the series converges to the distribution $\sum_{n=0}^\infty e^{-i\tau\omega n}=\frac{1}{1-e^{-i\tau\omega}}+\pi\sum_k\delta(\tau \omega+2\pi k)$. If we assume $\tau\omega \neq 2\pi k$ the delta term can be ignored. Notice that for finite $\N=\T/\epsilon$ and $\tilde\epsilon=0$, Eq.\ (\ref{eq:geomseries}) becomes the finite product 
\begin{equation}\medop{\prod}_{n=-\lfloor{\N/2}\rfloor}^{\lfloor{(\N-1)/2}\rfloor}
\Big[1-e^{i\tau (\omega_n-\omega )}\Big]^{-1}=\frac{e^{i\omega T/2}}{2i \sin(\omega \T/2)} \label{d4}\end{equation}
where the last expression holds for  $\tau=\epsilon$ and follows by expanding  $\frac{z^{\n}}{z^{\n}-1}$ in terms of the $\N$ roots of $1$, with $z=e^{-i\omega\epsilon}$.

The infinite product in (\ref{eq:geomseries}) indicates the inverse of $$\lim_{\n\to \infty}\medop{\prod}_{n=-\n}^{\n}\Big[1-e^{i\tau (\omega_n-\omega+i\tilde{\epsilon} \omega_n^2)}\Big]$$ as it follows e.g. by considering first $\T/\epsilon=2\N+1$ time steps in $\mathcal{H}$. 
We can split the product for finite $\N$ onto two terms with $n\geq 1$ and a $n=0$ contribution.
The convergence of the products with $n\geq 1$ 
is defined by the convergence of the series $\sum_{n=1}^\infty \exp[i\tau (\pm\omega_n-\omega+i\tilde{\epsilon} \omega_n^2)]$ which clearly converges absolutely for $\tau\tilde{\epsilon}>0$ ($|\exp[i\tau (\pm\omega_n-\omega+i\tilde{\epsilon} \omega_n^2)]|=\exp[-\tau\epsilon\omega_n^2]$).

In fact, if we let e.g. $\tilde{\epsilon}=\lambda \tau^2$ (with $\lambda>0$ a constant with units of [time]$^{-1}$) and $\tau \in \mathbb{C}$, the original infinite product defines an analytic function $F(\tau)$ in the subset of the complex plane defined by $\Re(\tau^3)>0$ (this can be proven by noting that the convergence is compactly normal in this region \cite{rem.13}).

We also ``conjecture'' that the limit $\tau\to 0^+$ of $F(\tau)$ takes the exact \emph{finite} value $\lim_{\tau\to 0^+}F(\tau)=e^{i\omega \T/2}/[2i \sin(\omega \T/2)]$, in agreement with (\ref{d4}), which we verified numerically.  This would imply for the corresponding action
\begin{equation}
    \lim_{\tau\to 0^+}{\rm Tr}[e^{i\mathcal{S}_\tau}]={\rm Tr}[e^{-iHT}] \,, \label{d5}
\end{equation}
 where we have restored the vacuum contribution $\mathcal{S}_\tau\to \mathcal{S}_\tau-\mathbbm{1}\T\omega/2$. We remark the difference with the usual continuum treatment which needs some regularization or 
an infinite constant encoded in the measure in order to provide a finite result (and which does not properly account the vacuum contribution in the CSPI case). Considering that for small $\tau$ we can relate this same trace with the PI expression of the partition function, the correctness of the conjecture would provide a rigorous continuum definition for this PI.

\section{\label{ApD}Stationary-action principle from a quantum mean value}
   The appearance of $S_{\rm cl}^{\star}$ in (\ref{eq:unitaryrel}) can also be understood by first noting that the average of the QA in the spacetime coherent states (Eq.\ (\ref{eq:coherentstcont})) is
\begin{equation}\label{eq:scla}
  \langle \mathcal{S}_\tau[j]\rangle_\alpha=  \frac{\langle \alpha(t)|\mathcal{S}_\tau[j]|\alpha(t)\rangle}{\langle \alpha(t)|\alpha(t)\rangle}=S_{\rm cl}+\gamma
\end{equation}
with $S_{\rm cl}$ the classical action along the trajectory $(q_{\rm cl}(t),p_{\rm cl}(t))$ defined as $q_{\rm cl}(t):=\sqrt{\tau}\frac{\alpha(t)+\alpha^\ast(t)}{\sqrt{2m\omega}}$, $p_{\rm cl}(t):=\sqrt{\tau}\frac{\alpha(t)-\alpha^\ast(t)}{i\sqrt{2  /m\omega}}$. Here $\gamma$ is a $j$ and $\tau$ independent constant arising from the vacuum energy.
The relation (\ref{eq:scla}) is a direct consequence of
\begin{equation}
A(t)|\alpha(t)\rangle=\alpha(t)|\alpha(t)\rangle
\end{equation}
 which also implies $\langle \sqrt{\tau} q(t)\rangle_\alpha =q_{\rm cl}(t)$ and  $\langle \sqrt{\tau} p(t)\rangle_\alpha =p_{\rm cl}(t)$.

As a consequence, the classical solution corresponds to a stationary value of the mean value (\ref{eq:scla}).  
This can be imposed directly in the Fourier basis $|\boldsymbol{\alpha}\rangle$ by noting that
$S_{cl}=\tau\sum_n(\omega_n-\omega)\alpha_{n}^*\alpha_n +\frac{j_n^*}{\sqrt{2\omega\tau}}\alpha_n+\frac{j_n}{\sqrt{2\omega\tau}}\alpha_{n}^*$.
Hence, in the present case the stationary condition
\begin{equation}\label{eq:statio}
  \frac{\partial }{\partial \alpha_n^\ast}\langle \mathcal{S}_\tau[j]\rangle_\alpha=0 \Leftrightarrow   \frac{\partial S_{\rm cl}}{\partial \alpha_n^\ast}=0\,
\end{equation}
  yields  $\alpha^\star_n=-\frac{j_n}{\sqrt{2\omega \tau}(\omega_n-\omega)}$ (the ``${}^\star$'' symbol indicates the solution). This defines the coherent trajectory solution $\alpha^\star(t)=\sum_n e^{-i\omega_n t}\alpha^\star_n/\sqrt{\T}$. Note that (\ref{eq:statio}) is equivalent to a variation on position and momentum in the Fourier basis, related to the previous position and momentum variables by a canonical transformation. In terms of the latter, the condition (\ref{eq:statio}) yields 
  \begin{equation}
      q^\star_{\rm cl}(t)=i\int dt'\, G(t-t')\tfrac{j(t')}{\sqrt{m}}\,
  \end{equation}
 and $p^\star_{\rm cl}(t)=mq^\star_{\rm cl}(t)$ with $(\partial_t^2+\omega^2)q_{\rm cl}^\star(t)=j(t)/\sqrt{m}$ in agreement with the Euler-Lagrange equation (and without any  $\tau$ dependence).
  
  The mean value of the QA along the classical solution is 
  \begin{equation}
      \langle {\cal S}_\tau[j]\rangle_{\alpha^\star}=
    -\tau\sum_n \frac{j_n j_{-n}}{2\omega\tau(\omega_n-\omega)}= S_{\rm cl}^{\star}\,,
  \end{equation}
  with $S_{\rm cl}^{\star}$ the classical action evaluated in the solution, in accordance with (\ref{eq:scla}), which is  
  $\tau$-independent as well. 
    Note also that $|\alpha^\star(t)\rangle=\mathcal{U}^\dag |\Omega\rangle$ is the {\it vacuum} of the shifted operators  ${\cal U}^\dag A_n {\cal U}$ with
    \begin{equation}\label{eq:utransl}
        \mathcal{U}:=e^{-\sum_n[\alpha^\star_n A^\dag_n-(\alpha_n^\star)^* A_n]}
    \end{equation}
such that $\mathcal{U}^{\dag} A_n \mathcal{U}=A_n-\alpha_n^\star$, in accordance with the definition above (\ref{eq:unitaryrel}). It is now clear that the constant factor arising from the action of $\mathcal{U}$ on $\mathcal{S}_\tau[0]$ must be $S_{\rm cl}^{\star}$: when evaluating the mean value of Eq.\ (\ref{eq:unitaryrel}) along the state $|\alpha^\star(t)\rangle$, the contribution of the second term vanishes since $\langle \alpha^\star(t)|\mathcal{U}^\dag\mathcal{S}_\tau[0]\mathcal{U}|\alpha^\star(t)\rangle=\langle \Omega|\mathcal{S}_\tau[0] |\Omega\rangle=0$.

Moreover, when one expands $S_{\rm cl}$ around the classical solution, the first order vanishes (Eq.\ (\ref{eq:statio})) while the second one is the same action without the source but evaluated along the ``fluctuating'' trajectory (larger orders of course vanish). In terms of quantum mean values this can be written as  $\langle \mathcal{S}_\tau[j]\rangle_\alpha=S_{\rm cl}^{\star}+\langle \mathcal{S}_\tau[0]\rangle_{\alpha-\alpha^\star}=S_{\rm cl}^{\star}+\langle \mathcal{U}^\dag \mathcal{S}_\tau[0]\mathcal{U}\rangle_{\alpha}$, which is just the expectation value of Eq.\ (\ref{eq:unitaryrel}).  
For small $\tau$, we can employ the previous considerations to reobtain (\ref{ZZ}) from familiar PI-like arguments: according to the discussion in sec.\ \ref{sec:consmalltau} we may write 
\begin{equation}
  Z_\tau[j]\approx \int \mathcal{D}^2\alpha(t) e^{i \langle \mathcal{S}_\tau[j]\rangle_\alpha}=e^{iS_{\rm cl}^{\star}}\int \mathcal{D}^2\alpha(t) e^{i \langle \mathcal{S}_\tau[0]\rangle_{\alpha}}\,, 
\end{equation}
 with the replacement $\langle \mathcal{S}_\tau[0]\rangle_{\alpha-\alpha^\star}\to \langle \mathcal{S}_\tau[0]\rangle_{\alpha}$ holding in the last equality because we are integrating over all trajectories. In the quotient  $Z_\tau[j]/Z_\tau[0]$ the ``fluctuation factor'' cancels out as in the conventional PI approach and in agreement with the more general $\tau$-independent derivation.


\end{document}